\documentclass[preprint,12pt]{elsarticle}

\usepackage{amsmath,amssymb,amsfonts}
\usepackage{algorithmic}
\usepackage{graphicx}
\usepackage{textcomp}
\usepackage{xcolor}
\usepackage{acronym}
\usepackage{verbatim}
\usepackage{multirow}
\usepackage{makecell}
\usepackage{todonotes}
\usepackage{balance}
\usepackage{url}
\usepackage{hyperref}
\usepackage{pifont}
\usepackage{array}

\newcommand{\sol}{\emph{PiNcH}}
\newcolumntype{P}[1]{>{\centering\arraybackslash}p{#1}}

\newcommand\blfootnote[1]{%
  \begingroup
  \renewcommand\thefootnote{}\footnote{#1}%
  \addtocounter{footnote}{-1}%
  \endgroup
}

\newcommand{\cmark}{\ding{51}}%
\newcommand{\xmark}{\ding{55}}%

\acrodef{UAV}{Unmanned Aerial Vehicle}
\acrodef{FPV}{First-Person View}
\acrodef{RSSI}{Received Signal Strength Indicator}
\acrodef{SDR}{Software Defined Radio}
\acrodef{CI}{Critical Infrastructure}
\acrodef{SR}{Success Rate}
\acrodef{FP}{False Positive}
\acrodef{TP}{True Positive}
\acrodef{FPR}{False Positive Rate}
\acrodef{TPR}{True Positive Rate}
\acrodef{FNR}{False Negative Rate}
\acrodef{TN}{True Negative}
\acrodef{FN}{False Negative}
\acrodef{ROC}{Receiver Operating Characteristic}
\acrodef{SDR}{Software Defined Radio}
\acrodef{RPAS}{Remotely Piloted Aircraft Systems}

\journal{Computer Networks}

\begin{document}

\begin{frontmatter}

\title{PiNcH: an Effective, Efficient, and Robust Solution to Drone Detection via Network Traffic Analysis}


\author{Savio Sciancalepore, Omar Adel Ibrahim, Gabriele Oligeri, \\ Roberto Di Pietro}

\address{Division of Information and Computing Technology\\ College of Science and Engineering, Hamad Bin Khalifa University \protect\\ Doha, Qatar \\ \{ssciancalepore, goligeri, rdipietro\}@hbku.edu.qa, oaibrahim@mail.hbku.edu.qa}


\begin{abstract}

\blfootnote{This is a personal copy of the authors. Not for redistribution. The final version of the paper is available through the ScienceDirect Digital Library, at the link: \url{https://www.sciencedirect.com/science/article/pii/S1389128619311764?via\%3Dihub}, with the DOI: \url{10.1016/j.comnet.2019.107044.}}We propose Picking a Needle in a Haystack (\sol), a methodology to detect the presence of a drone, its current status, and its movements by leveraging just the communication traffic exchanged between the drone and its Remote Controller (RC). \sol\ is built applying standard classification algorithms to the eavesdropped traffic, analyzing features such as packets inter-arrival time and size. \sol\ is fully passive and it requires just cheap and general-purpose hardware. To evaluate the effectiveness of our solution, we collected real communication traces originated by a drone running the widespread ArduCopter open-source firmware, currently mounted on-board of a wide range (30+) of commercial amateur drones. 
Then, we tested our solution against different publicly available wireless traces. The results prove that \sol\ can efficiently and effectively: (i) identify the presence of the drone in several heterogeneous scenarios; (ii) identify the current state of a powered-on drone, i.e., flying or lying on the ground;  (iii) discriminate the movements of the drone; and, finally, (iv) enjoy a reduced  upper bound on the time required to identify a drone with the requested level of assurance. The effectiveness of \sol\ has been also evaluated in the presence of both heavy packet loss and evasion attacks. In this latter case,  the adversary modifies on purpose the profile of the traffic of the drone-RC link to avoid the detection. In both the cited cases, \sol\ continues enjoying a remarkable performance. Further, the comparison against state of the art solution confirms the superior performance of \sol\ in several scenarios.
Note that all the drone-controller generated data traces have been released as open-source, to allow replicability and foster follow-up. 
Finally, the quality and viability of our solution, do prove that network traffic analysis can be successfully adopted for drone identification and status discrimination, and pave the way for future research in the area.

\end{abstract}


\end{frontmatter}

\section{Introduction}

\acp{UAV}, also known as drones, are becoming extremely popular due to their increasingly low prices and appealing functionalities. Indeed, drones are already adopted for several tasks such as inspections, perimeter control, remote surveillance, and emergencies \cite{Altawy2016}.

Unfortunately, drones represent the classical dual-use technology that, while providing great benefits, could also  be adopted for malicious intents, such as 
taking video/image pictures of---or violating \cite{londraiarport}---restricted-access areas \cite{Nassi_2019}, or even being used-as/carrying weapons against selected targets. The latter one is one of the major threats, not only for people \cite{maduro} but also for critical infrastructures such as airports and industrial sites, to name a few. 
The International Air Transportation Association (IATA) warned of ``an exponential increase in reports of Remotely Piloted Aircraft Systems (RPA) operating dangerously close to manned aircraft and airports" \cite{iata}.
While self-operated drones represent an expensive attack vector, also subject to GPS spoofing/jamming countermeasures, 
\ac{RPAS} drones are cheap and can be piloted for kilometers away from the operator due to the presence of a First Person View (FPV) channel. Furthermore, several unintentional near-hit and collisions have been reported between aircraft and drones \cite{asi}, due to the lack of understanding on the drones operators' side. Indeed, the frequency of these accidental events can only increase, due to the widespread use of drones for both recreational and commercial purposes \cite{amazon_drone}.

Moreover, drones can also be used to intentionally launch attacks against targets. For instance, an attack was launched in Syria on Russian military bases via a fleet of crudely made drones, each one equipped with GPS and powered by what appeared to be lawn mower engines, with each drone carrying nearly half a kilogram of high potential explosives \cite{asi}.

Several drone countermeasures have been developed and already deployed. Some of them involve the use of jammers to disable the remote controller of the drone, hence forcing it to land, the use of other drones to chase the not-authorized one and, finally, weapons to shoot the drone down \cite{Lin2018}. While several start-ups have already developed different anti-drone solutions mainly based on radar detection and jamming, an interesting academic solution is represented by ADS-ZJU \cite{shi}, where the authors propose an integrated approach combining multiple passive surveillance technologies to realize drone detection, localization, and radiofrequency jamming. \textcolor{black}{At the same time, several projects on drone detection have been funded by the European Union within the H2020 program, including \emph{SafeShore}\footnote{http://safeshore.eu} and \emph{Aladdin}\footnote{https://aladdin2020.eu}, to name a few (see Sec.~\ref{sec:related}) for more details.}

While drone counter-measures have already reached a significant level of reliability, drone detection can only rely on a few effective techniques \cite{azari}. Among the various techniques, four major strategies are emerging: (i) visual detection; (ii) audio detection; (iii) radar; and, (iv) RF detection. Visual detection mainly relies on the distribution of camera equipment in the area to be protected and the implementation of video processing techniques to identify anomalies in the video stream \cite{rozantsev}. Audio detection resorts to the generation of an audio signature of the drone propellers to be used to train a classifier \cite{kim}; 
such a technique further requires arrays of microphones to be deployed in the area to be monitored \cite{Busset2015}. Conversely, radar detection involves the transmission of RF signals to receive an RF echo that can be identified and tracked \cite{hoffmann}. Radars, while being the most powerful among the detection strategies, are expensive equipment that eventually might not be effective to detect small devices such as drones---since the radar signature is quite blurred. 
Finally, RF-based techniques resort to the generation of RF fingerprints by looking at the communication channel between the drone and its remote controller \cite{nguyen}. RF fingerprinting is a promising technique that has been used for several purposes, but it requires specific equipment such as \acp{SDR}. However, cheap SDRs available on the market, such as the RTL-SDR, cannot be considered fully reliable, especially when operating at high frequencies.

Thus, we observe that the current literature still misses a viable and cheap solution, not requiring any dedicated hardware, while enabling the recognition not only of the presence of an \ac{RPAS} drone, but also its current state in a real-time perspective. What is more, such a solution should be robust to packet loss, as well as evasion attacks.

{\bf Our contribution.} In this manuscript, we present \emph{Picking a Needle in a Haystack} (\sol), a solution to detect an RPAS drone based on encrypted network traffic analysis. Compared to our initial contribution in \cite{Sciancalepore2019_WiseML}, \sol significantly improves the current state of the art in remotely-controlled drones detection with several contributions:
\begin{itemize}
    \item \emph{Drone detection.} \sol\ can detect the presence of a drone in different heterogeneous scenarios such as a library, a cafeteria, a conference, and outdoor areas.
    \item \emph{Drone state identification.} \sol\ can discriminate the current state of a powered-on drone, i.e., if the drone is either flying or lying on the ground.
    \item \emph{Detection delay.} We provide a statistical analysis of the detection delay for each of the aforementioned classification scenarios as a function of the requested level of assurance.
    \item \emph{Packet Loss}. We study the effect of packet loss on the performance of \sol, by testing the detection accuracy when the eavesdropping equipment is located at increasing distances from the drone. We demonstrate \sol\ to be able to overcome packet loss issues and to guarantee very high detection accuracy even at 200 meters from the position of the drone.
    \item \emph{Evasion strategy}. We test the effectiveness of \sol\ in the presence of evasion strategies, where the adversary reshapes the traffic profile of the controller-drone communication channel by introducing random delays between the message packets. We show that the effectiveness of such a strategy strongly depends on the specific deployment scenario, though, in general, it should be noted that it significantly reduces the maneuverability of the drone.
    \item \emph{Movement identification.} \sol\ can infer on the current movement of the drone, discriminating if the drone is increasing its altitude, moving forward, backward, left or right.
\end{itemize}

To the best of our knowledge, \sol\ represents the first comprehensive solution able to detect an RPAS drone and, at the same time, its current state in real-time, looking only at the wireless traffic. Other unique features enjoyed by \sol\ are its robustness to packet loss and the possibility to reject evasion attacks, based on the specific deployment scenario. 

\textcolor{black}{At the same time, we stress that \sol\ aims to detect an RPAS drone, where a remote controller is communicating with the drone. Indeed, \sol\ cannot detect an autonomous drone, given that no communication between the controller and the drone is involved. However, for these scenarios, other solutions based on acoustic, visual, or radar techniques are applicable. In addition, due to its lightweight and non-invasive features, \sol\ can be integrated with other drone detection techniques, to build a multi-method framework leveraging multiple means to detect an approaching drone.}

The results included in this paper have been obtained by using the popular drones' firmware ArduCopter, within the Ardupilot operating system. Thus, consistently with other recent work in the literature such as \cite{Chen_cns2019}, beyond the 3DR SOLO drone used in this paper, our results are fully applicable also to over 30 products, including DJI and HobbyKing vehicles, to name a few\footnote{\url{http://ardupilot.org/copter/docs/common-rtf.html\#common-rtf}}\footnote{\url{http://ardupilot.org/copter/docs/common-autopilots.html}}.

The drone-controller data traces we have generated in this work have been released as open-source (available at \cite{dataset}), to allow practitioners, industries, and academia to verify our claims and to use them as a basis for further development.

Finally, we remark that the aim of this paper is neither to propose a new intrusion detection algorithm nor to discuss new machine learning techniques. Indeed, we discuss and demonstrate, through an extensive measurement campaign run over an open-source operating system, that the presence, the status, and the specific operational mode of commercial remotely-operated drones can be identified using already available classification tools, paving the way to further research efforts by both critical infrastructure defense teams and drone operators.

{\bf Paper organization.} The paper is organized as follows: Section \ref{sec:related} reviews related work, Section \ref{sec:system_and_adversary_model} introduces the system and the adversary models assumed in this work, while Section \ref{sec:scenario_considerations} details the measurement scenario and provides some details about the measurements and the characterization of the network traffic generated by the drone. Section \ref{sec:classification_methodology} introduces the methodology we used for the acquisition, processing, and classification of the network traffic generated by the drone and the remote controller, while Sections \ref{sec:scenario_identification} and \ref{sec:detecting_a_drone} show the performance of our proposal for detecting the state of the drone and the presence of the drone in different heterogeneous scenarios, with a look also on the detection delay. Section \ref{sec:movement_identification} introduces the results related to the identification of each movement the drone can take, while Section \ref{sec:robustness} reports the performance of \sol\ at increasing distances from the drone and when evasion strategies are applied by the adversary. Finally, Section \ref{sec:conclusions} reports some concluding remarks.

\section{Related work}
\label{sec:related}

In the last years, the widespread diffusion of commercial drones has paved the way for several research contributions discussing the potential identification of \acp{UAV} in a certain area of interest.

The authors in \cite{Nassi_2019} built a proof-of-concept system for counter-surveillance against spy drones by determining whether a certain person or object is under aerial surveillance. They show methods that leverage physical stimuli to detect whether the drone’s camera is directed towards a target in real-time. They demonstrate how an interceptor can perform a side-channel attack to detect whether a target is being streamed by analyzing the encrypted \ac{FPV} channel that is transmitted from a real drone (DJI Mavic) in two use cases: when the target is a private house and when the target is a subject.  A similar target, i.e. video streaming detection, has been investigated in the recent work by the authors in \cite{birnbach2017wi}, focusing on the signal strength in the communication between the drone and its controller. Although being a significant step towards drone identification, these solutions are specifically designed to identify drones that are employed to target a specific target, while not being suitable for drone's detection at large or for drones that do not necessarily feature FPV.

The authors in \cite{Shoufan2018} showed that the radio control signal sent to a UAV using a typical transmitter can be captured and analyzed to identify the controlling pilot using machine learning techniques. The authors collected the messages exchanged between the drone and the remote controller, and used them to train multiple classifiers. They observed that the best performance is reached by a random forest classifier achieving an accuracy of around 90\% using simple time-domain features. The authors also provided extensive tests showing that the classification accuracy depends on the flight trajectory. In addition, they proved that the control signals, i.e., pitch, roll, yaw, and throttle, have specific importance for pilot identification. The work focused on a scenario where civil \acp{UAV} are remotely controlled by different pilots, there is no (or weak) authentication on the ground-to-aircraft command channel, and also there is little to null difference in the low-level timing or power of the control signals. In addition, they assumed that the pilots could carry out identical maneuvers, as well as the existence and availability of trustworthy recordings of each pilot’s behavior. 
While exploiting the same principle, i.e., classification of the traffic, the related work focuses on the pilot and not on drone identification.

The authors in \cite{Nguyen2017} explored the feasibility of RF-based detection of drones by looking at radio physical characteristics of the communication channel when the drones' body is affected by vibration and body shifting. The analysis considered whether the received drone signals are uniquely differentiated from other mobile wireless phenomena such as cars equipped with Wi-Fi or humans carrying a mobile phone. The sensitivity of detection at distances of hundreds of meters as well as the accuracy of the overall detection system are evaluated using a \ac{SDR} implementation. Being based on both \ac{RSSI} and phase of the signals, the precision of the approach varies with the distance of the receiver from the transmitter. In addition, the solution resorts to physical layer information and special hardware (SDR), while our current contribution only exploits network layer information that can be collected by any WiFi device.

An identification mechanism based on the correlation between motion observed from an external camera and acceleration measured on each UAV's accelerometer is proposed by the authors in \cite{ruiz}. This solution combines FPV information with accelerometer information to remotely control a subset of swarm drones that are not provided with a camera, and therefore it requires the collaboration of one or more drones in the swarm to perform the identification.

Fingerprinting of wireless radio traffic at the network layer is emerging as a promising technique to uniquely identify devices in the wild. The authors in \cite{xu} proved that the extraction of unique fingerprints provides a reliable and robust means for device identification. 

A fingerprinting approach for drone identification is proposed in \cite{li}. The authors analyzed the WiFi communication protocol used by drones and developed three unique methods to identify a specific drone model: (i) examining the time intervals between probe request frames; (ii) utilizing the signal strength carried in the frame header; and, finally (iii) exploiting some frame header fields with specific values. However, fingerprint approaches require specific equipment to be used, such as the Software Defined Radios (SDRs).

Network-Based traffic classification is proposed in \cite{bisio}. The authors describe a WiFi-based approach aimed at detecting nearby aerial or terrestrial devices by performing statistical fingerprint analysis on wireless traffic. They proved network-layer classification to be a viable means to classify classes of drones, i.e., aerial, terrestrial, and hybrid scenarios. However,  their approach does not take into account the identification of drone traffic compared to standard WiFi traffic. The same authors extend the aforementioned contribution in \cite{Bisio2018_TVT}, by proposing a WiFi statistical fingerprint method to drone detection. Their solution can identify the presence of a drone and the associated video streaming.
Similarly, the authors in \cite{Alipour2019} adopted encrypted traffic analysis techniques to identify the presence of a drone, considering Parrot, DJI, and DBPower drones. 
However, these solutions (available on the arXiv portal later than the present contribution) take into account very specific drones based on a proprietary architecture (DJI, Parrot), and they can detect only the presence of the drone without inferring its current status. Finally, the authors prove the feasibility of their solution considering self-generated traces as network noise, while in this work, we consider 5 publicly available data-sets and one (generated by us) from an outdoor area.

Another passive detection technique is proposed in \cite{fu}. The authors presented a technique specifically designed for two reference scenarios: (i) a drone communicating with the ground controller, where the cyclo-stationarity signature of the drone signal and pseudo Doppler principle are employed; and, (ii) a drone that is not sending any signal, where a micro-Doppler signature generated by the RF signal is exploited for detection and identification. Also in this case, the authors resort to both SDRs and physical layer fingerprinting, thus making their solution very hardware-invasive.

Machine learning techniques have been successfully used for other purposes in this research field. In \cite{Jeong2017}, the authors proposed a wireless power transfer system that predicts the drone's behavior based on the flight data, utilizing machine learning techniques and Naive Bayes algorithms.

In \cite{Park2016}, the authors demonstrated that machine learning can successfully predict the transmission patterns in a drone network. The packet transmission rates of a communication network with twenty drones were simulated, and results were used to train the linear regression and Support Vector Machine with Quadratic Kernel (SVM-QK).

Standard anti-drone active detection techniques resort to radar \cite{lee, quilter, roding}. Nevertheless, those techniques involve the transmission of signals and specific devices for the detection of the echo fingerprint. 

The authors in \cite{jain} analyze the basic architecture of a drone and propose a generic drone forensic model that would improve the digital investigation process. They also provide recommendations on how one should perform forensics on the various components of a drone such as a camera and Wi-Fi.

\textcolor{black}{
Finally, we highlight that several projects tackling the detection of amateur or remotely-piloted drones have been funded by the European Union within the H2020 program, including \emph{SafeShore}\footnote{http://safeshore.eu} and \emph{Aladdin}\footnote{https://aladdin2020.eu}, to name a few (see Sec.~\ref{sec:related}) for more details. With specific reference to coastal border-surveillance, the SafeShore project aims at detecting Remotely Piloted Aircraft Systems (RPAS) carrying out illegal activities via passive and low-cost technologies, being very close to the aim of this project. The threats and the system requirements considered in this project has been discussed by the authors in~\cite{buric2017_MTA}. The project SafeShore has also launched a specific challenge, namely the \emph{drone-vs-bird detection challenge}, dedicated to addressing one of the many technical issues arising in this context \cite{coluccia2017_avss}, \cite{coluccia2019_avss}.
Another solution is the one that is being developed within another EU project, namely Advanced hoListic Adverse Drone Detection, Identification \& Neutralization (ALADDIN). The ALADDIN project aims to develop a complete product for the drone detection problem, leveraging a combination of the systems described above, including radar, video, sound, and further detection methods (for more details and publication overview, see the website\footnote{https://aladdin2020.eu/reports-publications/} of the project).
}

To sum up, none of the previous contributions can detect a drone and its current status by only exploiting the wireless traffic. Moreover, differently from the current literature, our contribution provides a thorough measurement campaign adopting a widely accepted open-source operating system and firmware for drones, included in over 30 products, as well as a detailed analysis adopting a different type of wireless network traffic to prove its robustness.
Finally, a study of the effectiveness of evasion attacks against drone detection strategies based on traffic analysis is provided in our contribution, as well as how packet loss phenomena can affect our strategy, differently from the previous work. 

We will provide a qualitative and experimental comparison between \sol\ and related approaches in Section \ref{sec:comparison}.

\section{System and Adversary Model}
\label{sec:system_and_adversary_model}
{\bf Adversarial model.} In this paper, we assume a scenario characterized by an RPAS drone flying over a no-fly-zone, where the GPS is not available. Indeed, we recall that drones leveraging GPS navigation can be easily defeated by adopting GPS-spoofing and jamming techniques. We assume the drone is remotely controlled by a malicious operator that intentionally wants to fly the drone across the border of a restricted-access area such as an airport, industrial plant, or critical infrastructure. We also assume that the adversary is deploying additional countermeasures preventing drone identification, such as dynamically changing the MAC addresses of the network interfaces of both the drone and the remote controller. In addition, we assume the link between the RC and the drone is encrypted at the layer-2 of the communication link, and therefore packet content cannot be inspected. We assume also that the link between the controller and the drone cannot be jammed, as the same frequencies are used for legitimate communications by other devices.

Finally, we initially assume that the adversary does not apply any evasion techniques, e.g., it does not modify the transmission rate and the length of the packets to mitigate the detection. However, in Section \ref{sec:robustness} we discuss the effectiveness of \sol\ in presence of a particular type of evasion attack, where the adversary delays on purpose the packets to be transmitted by the RC and the drone, to avoid the detection. Overall, while implementing such countermeasures might improve the probability of escaping detection, their overall efficacy is not guaranteed, as a new training of our model would suffice to identify again the status of the drone. Further, the application of evasion strategies requires further feasibility studies based on the specific requirements of the application \cite{Zhang2016adversarial}. Indeed, introducing an artificial delay in the communication between the RC and the drone could affect user experience, to the point that the planned mission could be aborted. In addition, none of the actual commercial products implement such features, making our proposed solution effective.

{\bf System model.} Our main goal is the passive detection of the remotely-controlled drone without resorting to: (i) active radar technology \cite{quilter}; (ii) audio/visual detection \cite{saqib}; or, (iii) physical stimuli to FPV \cite{Nassi_2019}. Our solution does not require any intervention in the already existing ICT infrastructure and it does not conflict with any already deployed RF system. Indeed, \sol\ exploits only the messages transmitted between the remote controller and the drone, and therefore it only requires a fully passive eavesdropper to be deployed in the region to be controlled.

\begin{figure}[htbp]
\includegraphics[width=.8\columnwidth]{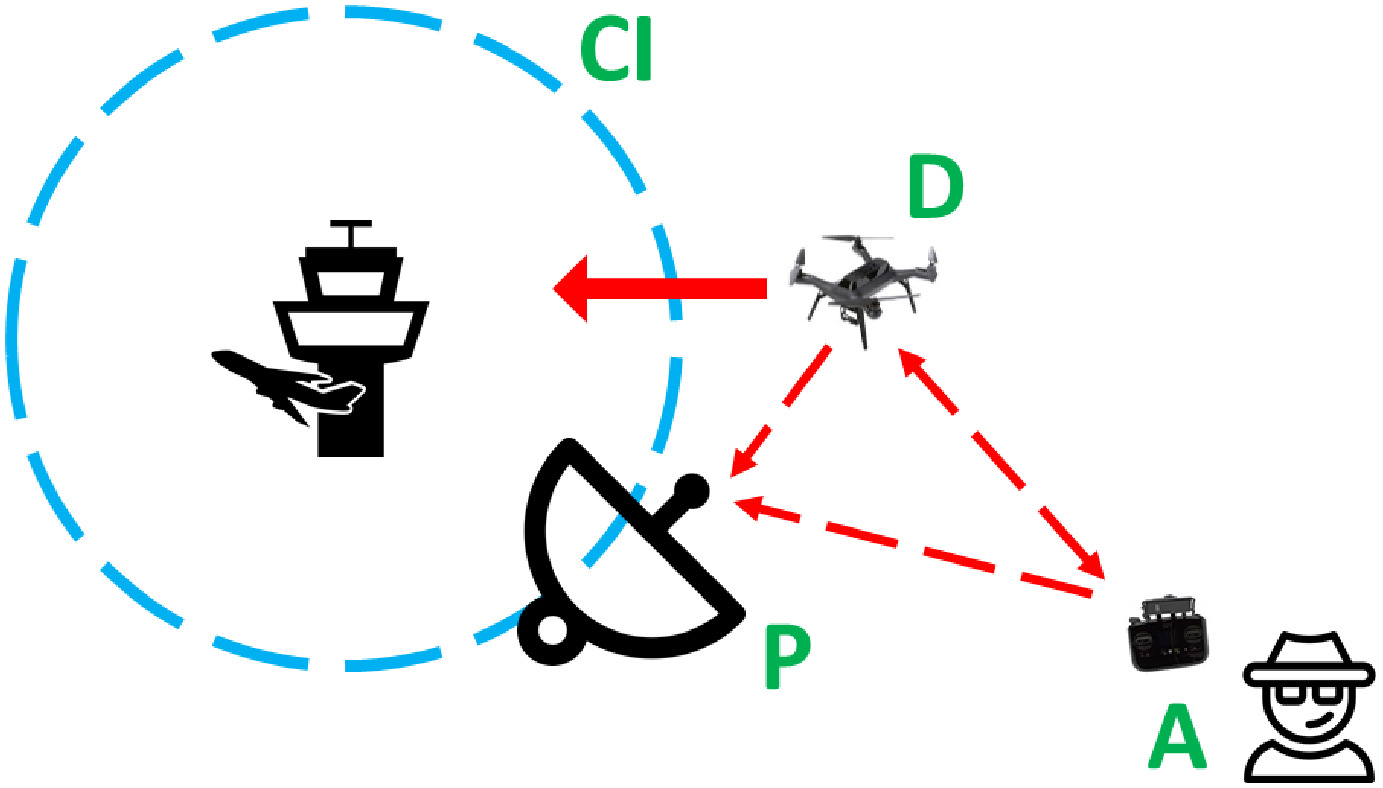}
\centering
\caption{System model: a no-fly zone (e.g. critical infrastructure ---  CI) featuring a WiFi probe (P) to detect an approaching drone (D) remotely controlled by an adversary (A).}.
\label{fig:model}
\end{figure}

Figure \ref{fig:model} wraps up on the system and adversarial models: the adversary (A) is determined to remotely fly a drone (D) into a no-fly zone (CI). Our solution can detect the drone's presence by simply deploying a WiFi probe (P). We observe that the WiFi probe can eavesdrop both the traffic from the controller to the drone (A-D) and the one from the drone to the controller (D-A).

Despite our measurement campaign was performed using the 2.4 GHz frequency band and the WiFi communication technology, the methodology proposed in this paper can be easily extended to work on any operating frequency, provided that one (or more) receivers are available on that particular frequency, independently on the usage of any encryption technique. Indeed, we remark that most of the commercial drones, including the ones running the ArduCopter Operating System, use either the 2.4~GHz frequency band or the 900~MHz ISM band.

\section{Measurement scenario and Preliminary considerations}
\label{sec:scenario_considerations}

\begin{figure}[htbp]
\includegraphics[width=0.8\columnwidth]{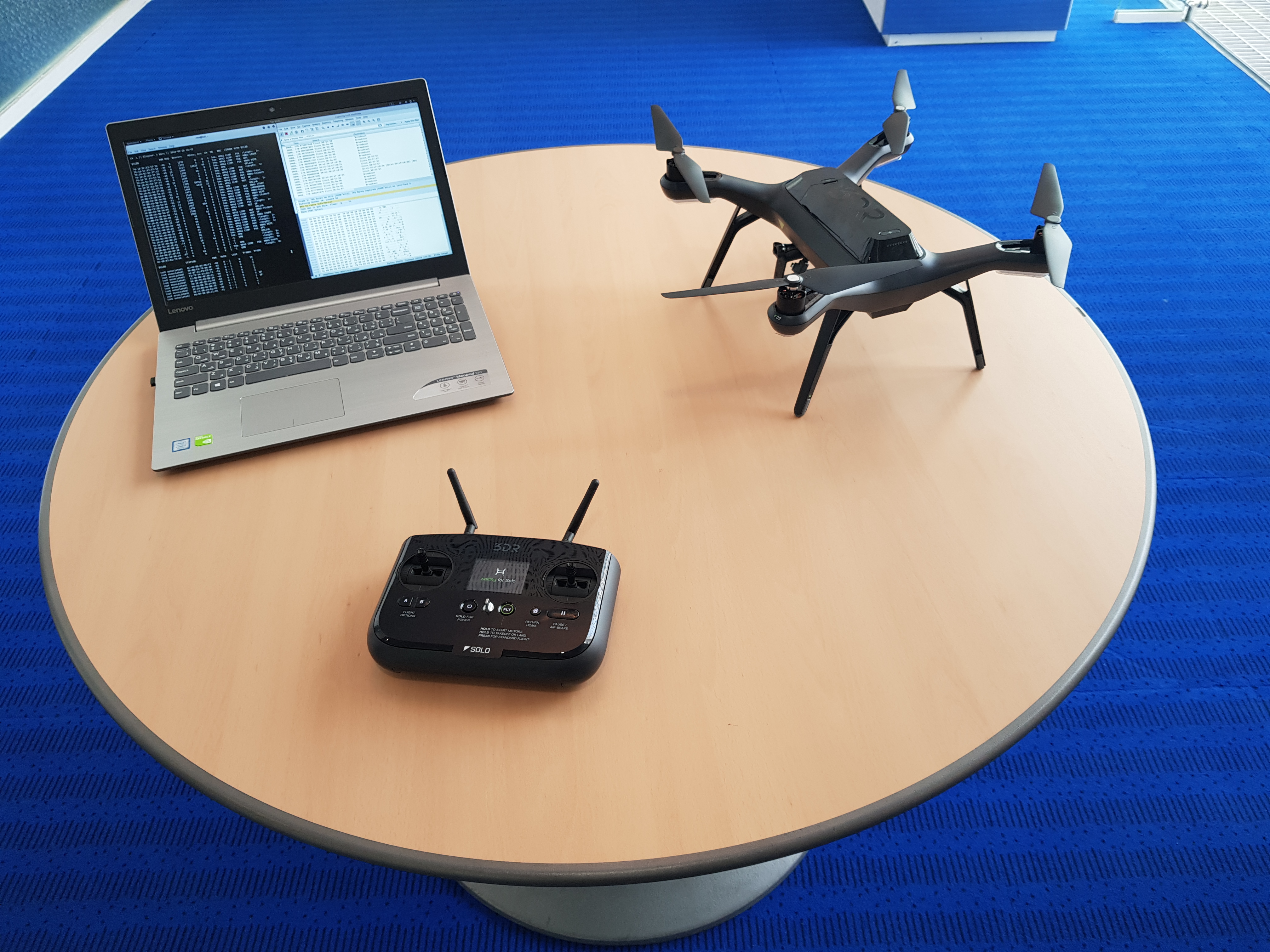}
\centering
\caption{Our measurement set-up: the drone, the remote controller, and the laptop we used to eavesdrop the radio spectrum.}.
\label{fig:setup}
\end{figure}

Our measurement scenario is constituted by a 3DR SOLO drone\cite{3dr} and a wireless probe capable of eavesdropping the radio traffic. The 3DR Solo drone is an open-source architecture featuring the Pixhawk 2.0 flight controller and the ArduCopter 3.3 firmware. The drone has been configured for the \emph{manual mode}, i.e., GPS switched off, and therefore, being able to fly both in indoor and outdoor environments. As a wireless probe, we adopted Wireshark 2.4.5, running in a Lenovo Ideapad 320 featuring Linux Kali 4.15.0. We configured the WiFi card of our laptop to work in monitor mode, eavesdropping and logging all the transmitted packets by either the remote controller or the drone. 
Figure \ref{fig:setup} shows our measurement set-up.

Subsequently, we collected several packets from the controller-drone link, while the drone was performing two different types of actions, as depicted below:
\begin{itemize}
    \item \emph{Steady} (S). The drone is associated with the remote controller but it lays on the ground.
    \item \emph{Movement} (M). The drone is flying around, performing several movements in all directions.
\end{itemize}
Table \ref{tab:link_stats} wraps up on the two states of the drone and provides ground for some preliminary considerations on the collected measurements. Firstly, we break down the communication link into 4 different flows: (i) the packets sent by the RC to the Drone; (ii) the packets sent by the Drone to the RC; (iii) the packets broadcast by the RC; and, finally (iv) the packets broadcast by the drone. 
Secondly, we considered a measurement lasting for about 10 minutes for the Steady-state, where the drone is associated to the remote controller but it is lying on the ground; then, we unscrewed the propellers from the drone and we ``flew" the drone for about 10 minutes for the Movement state. 
During the flight, we performed different maneuvers by continuously moving the control sticks. The percentage of exchanged packets is similar in both the drone's states, i.e., about 36\% of the traffic is transmitted by the RC, while about 58\% of the traffic is received by the RC. This is consistent, as the drone is required to transmit more traffic to the RC, for the RC to be always able to know precisely the full status of the drone. Finally, we observe the presence of broadcast traffic transmitted by the RC (about 6\%), while we did not detect any broadcast communication transmitted by the drone after the pairing process.

\begin{table}[htbp]
\centering
\caption{Drone's states and flows: We considered 2 different states, i.e., Steady and Movement, and 4 different unidirectional communication flows, i.e., RC to Drone, Drone to RC, RC to broadcast and, finally, Drone to Broadcast.}
\begin{tabular}{|c|l|l|c|c|}
\hline
\multicolumn{1}{|l|}{\textbf{State}}             & \textbf{Source} & \textbf{Destination} & \textbf{N. of Pkts} & \textbf{Flow/Link (\%)} \\ \hline
\multicolumn{1}{|c|}{\multirow{4}{*}{\makecell[c]{Steady \\ (S)}}} & RC              & Drone                & 32706                    & 35.8                \\ 
\multicolumn{1}{|l|}{}                              & Drone           & RC                   & 52856                    & 57.8                \\ 
\multicolumn{1}{|l|}{}                              & RC              & Broadcast            & 5789                     & 6.4                 \\ 
\multicolumn{1}{|c|}{}                              & Drone           & Broadcast            & 0                        & 0                   \\ \hline
\multirow{4}{*}{\makecell[c]{Movement \\ (M)}}                           & RC              & Drone                & 32868                    & 34.6                \\ 
                                                    & Drone           & RC                   & 56248                    & 59.2                \\ 
                                                    & RC              & Broadcast            & 5837                     & 6.2                 \\ 
                                                    & Drone           & Broadcast            & 0                        & 0                   \\ \hline
\end{tabular}
\label{tab:link_stats}
\end{table}

\begin{figure}[htbp!]
\includegraphics[width=.75\columnwidth]{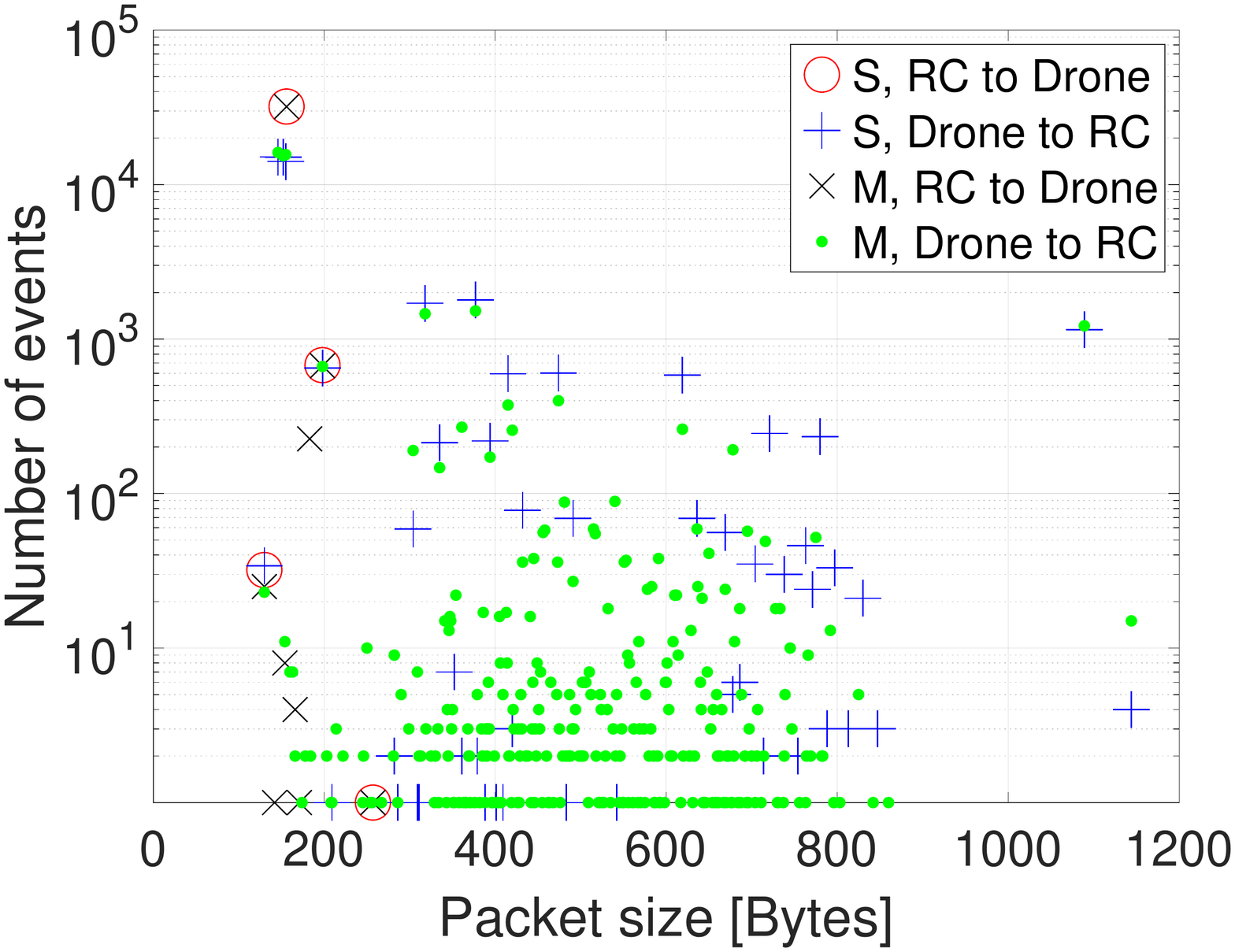}
\centering
\caption{Packet size analysis: Number of packets as a function of their size considering the four communication flows. The total number of collected packets (events) is 174677.}
\label{fig:packet_size_analysis}
\end{figure}

\begin{figure}[htbp!]
\includegraphics[width=.75\columnwidth]{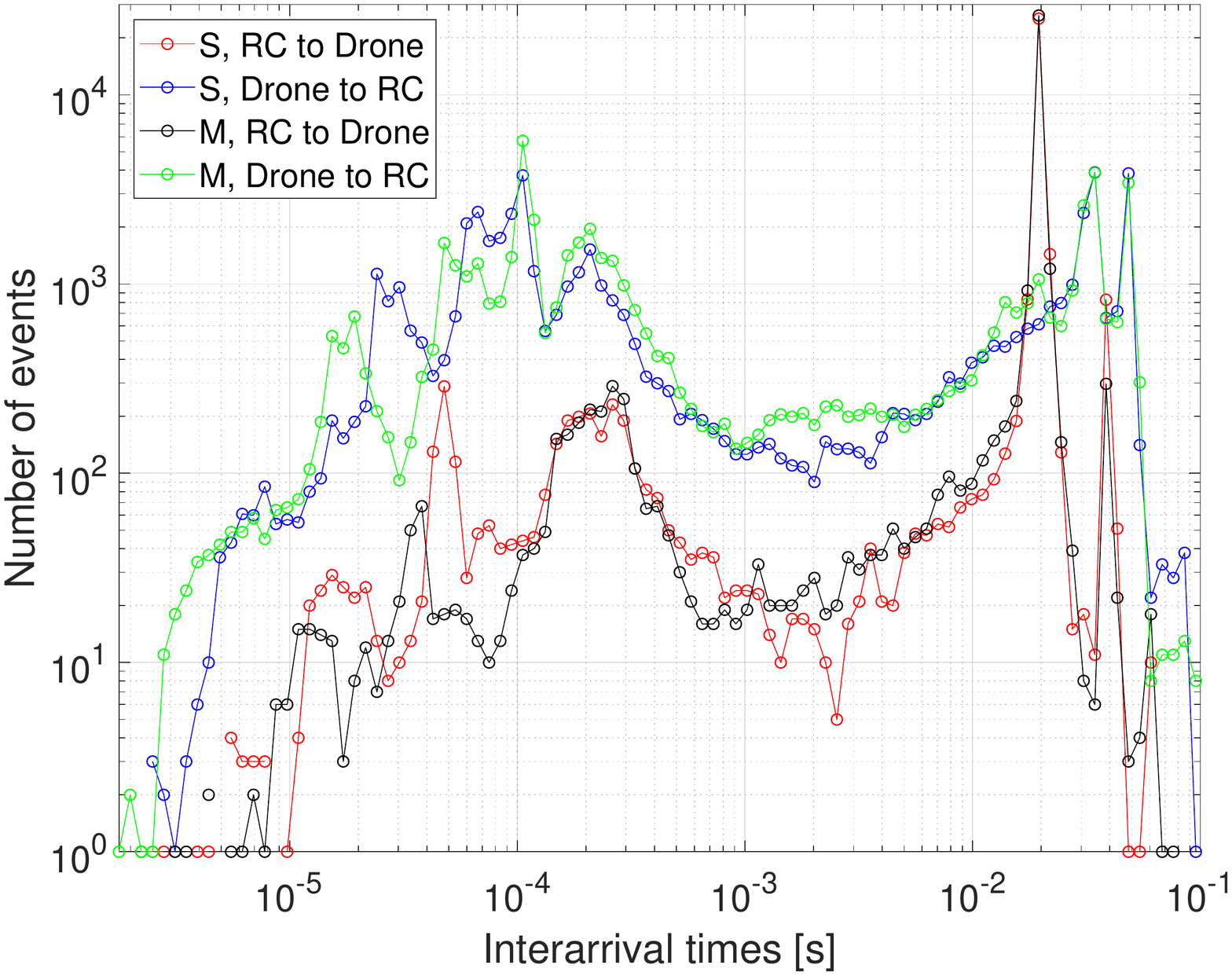}
\centering
\caption{Packet size analysis: Number of packets as a function of their size considering the four communication flows. The total number of collected packets (events) is 174677.}
\label{fig:time_analysis}
\end{figure}

%
In the following analysis, we focus on two packet attributes: \emph{packet size} and \emph{packet inter-arrival time}. For both the attributes, we took the previously identified flows from Section \ref{sec:scenario_considerations}, and we analyzed the size of the packets for the 4 different flows combining the two states, i.e., Steady and Movement, and the packets traveling from the RC to the drone, and from the drone back to the RC. 
\\
{\bf Packet size analysis.} Figure \ref{fig:packet_size_analysis} shows the frequency distribution function associated with each packet size belonging to the four different flows. Firstly, the vast majority of packets transmitted in the configuration \emph{S, RC to drone} (red circles) have the size equal to 156 Bytes. A similar phenomenon can be observed for \emph{M, RC to drone} (black crosses). Conversely, the flows coming from the drone are characterized by very different packet sizes spanning from 130 to 1144 Bytes, while more packets are transmitted by the drone when it is in the Steady-state, i.e., blue crosses have on average higher values than the green stars. These considerations motivated us to consider as ``discriminating features'' also the mean and the standard deviation of the packet size, computed over time windows of a different duration.
\\
{\bf Inter-arrival time analysis.} We extracted the time associated with all the events belonging to the same flow and we differentiated them obtaining the inter-arrival times. Figure \ref{fig:time_analysis} shows the number of packets as a function of their inter-arrival time. Firstly, we observe periodic packets at $20 ms$ and $40 ms$ transmitted by the RC to the drone in both the drone's states. Then, we observe how the two drone's states, i.e., Steady and Movement, are characterized by almost the same profile: there are only minor differences at about $31 \mu s$ and between $50 \mu s$ and $90 \mu s$. Indeed, the correlation coefficients computed over the frequency distribution functions are 0.71 and 0.75, for the flows transmitted and received by the remote controller, respectively. Finally, for both the considered drone's state, we observe that the drone is transmitting more data then the RC, i.e., blue and green curves are higher than the black and red ones. 
\\
{\bf Broadcast traffic.} Packets transmitted to the broadcast address by the remote controller have the same size (289 bytes) and an interarrival time equal to $100ms$ in both the drone's state. Given their strictly periodic nature, we do not consider them further in our analysis.
\section{Methodology}
\label{sec:classification_methodology}
We now introduce \sol, our solution to drone's detection, and we test it against the previously introduced drone's states, i.e., Steady and Movement. As our classification tools, we adopted a series of scripts developed using Matlab R2019a and the \emph{Machine Learning} toolbox. For our analysis, we consider the following configuration:
\begin{itemize}
    \item \emph{Classifiers.} We considered the Random Forest algorithm, being the best among those we tried (see \cite{Sciancalepore2019_WiseML} for more details).
    \item \emph{Flows.} We consider 3 different flows, i.e., RC to drone, drone to RC, and the overall link.
    \item \emph{Features.} We look at 6 different features, i.e., interarrival time, packet size, mean and standard deviation computed over a certain number of samples of interarrival time and packet size.
\end{itemize}

\textcolor{black}{
Figure \ref{fig:processing} introduces the details of \sol, i.e., our proposed solution.\\
\begin{figure*}[htbp!]
\includegraphics[width=\columnwidth]{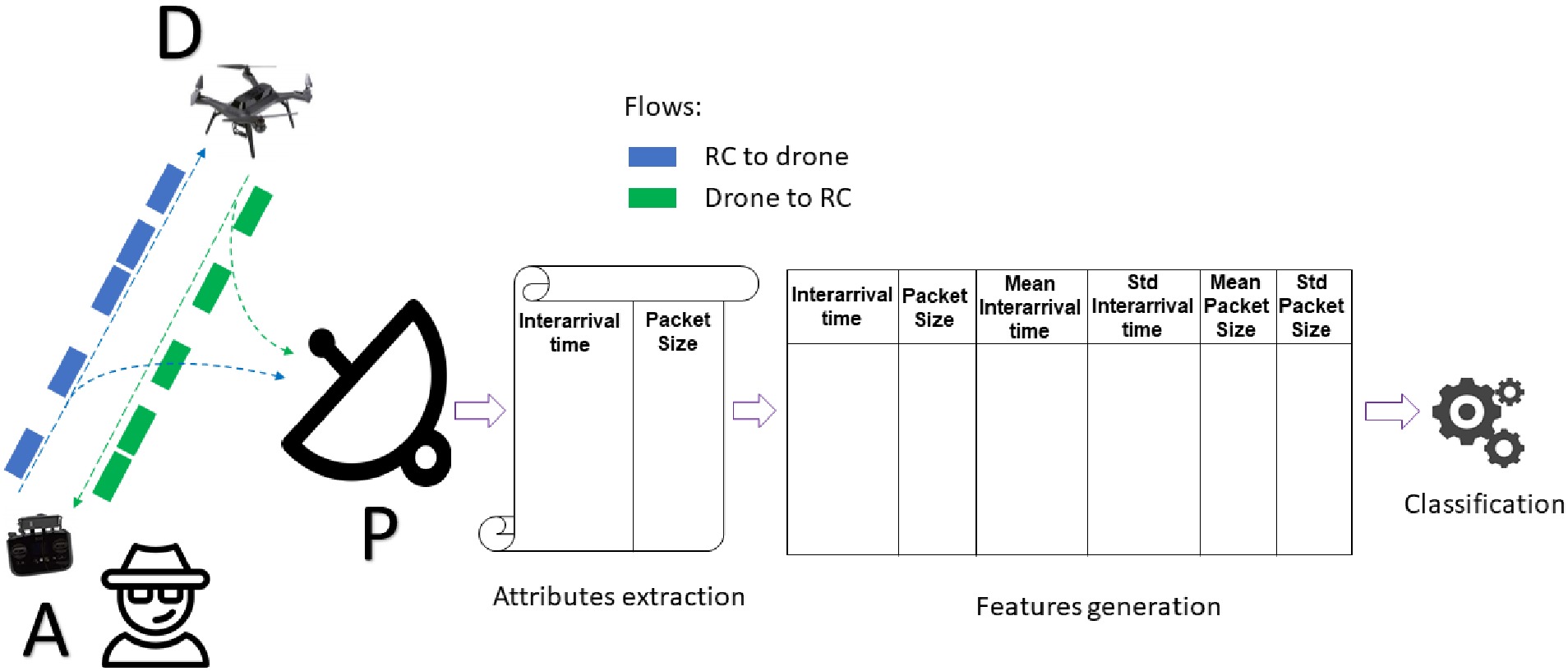}
\centering
\caption{\sol\ classification methodology: (i) WiFi radio eavesdropping, (ii) attribute extraction, (iii) feature generation, and (iv) classification.}
\label{fig:processing}
\end{figure*}
The notation used throughout the rest of this work is summarized in Tab.~\ref{tab:notation}. Lower case letters refer to a scalar value, while a boldface lowercase letter, e.g., $\mathbf{l}$, is used to represent a vector. The vector elements are listed within brackets.  
\begin{table}[htbp]
\centering
\color{black}
\caption{\textcolor{black}{Notation used throughout the paper}.}
\label{tab:notation}
\begin{tabular}{|P{0.13\columnwidth}|P{0.87\columnwidth}|}
\hline
{\bf Notation} & {\bf Description} \\
\hline
$N$ & Number of Packets of dataset $A$. \\
\hline
$a_n$ & Arrival Time of the generic n-th packet. \\
\hline
$p_n$ & Size of the generic n-th packet. \\
\hline
$s_n$ & Source MAC address of the generic n-th packet.\\
\hline
$d_n$ & Destination MAC address of the generic n-th packet.\\
\hline
$J$ & Number of Packets of dataset $B$.\\
\hline
$t_n$ & Interarrival Time between the n-th packet and $\left( n+1\right)$-th packet. \\
\hline
$F$ & Number of Features extracted from each trace. \\
\hline
$x_{n,f}$ & Generic f-th feature of the n-th packet. \\
\hline
$L$ & Number of labels. \\
\hline
$y_n$ & Label associated to the n-th packet, being $y_n = \left[ 0,1\right]$. \\
\hline
$W$ & Window size used for features computation. \\
\hline
$D$ & Overall matrix used for classification. \\
\hline
$M$ & Overall number of packets used for classification. \\
\hline
$\mathbf{x}$ & Feature set in the matrix $D$. \\
\hline
$\mathbf{y}$ & Label set in the matrix $D$. \\
\hline
$h_k (\mathbf{x})$ & Generic k-th decision tree classifier working on the features set $\mathbf{x}$. \\
\hline
$a_t$ & Threshold value on the node $t$ of the decision tree classifier. \\
\hline
$\Theta_k$ & Parameters set of the k-th decision tree classifier. \\
\hline
$H(T)$ & Entropy of the data-set $T$. \\
\hline
$P_l$ & Probability that the generic label $l$ is present in a dataset. \\
\hline
$IG(T,a)$ & Information Gain derived from the split of dataset $T$ with the threshold $a$. \\
\hline
$vals(x_k)$ & Set of possible values for the feature $x_k$. \\
\hline
\end{tabular}
\end{table}
We assume a communication link between the drone (D) and the adversary (A) constituted by two flows: remote controller to drone and backward. In summary, \sol\ requires that both the flows are eavesdropped by a WiFi probe, collecting two attributes: the interarrival time between subsequent packets and the packet size. Subsequently, a set of features is generated from the attributes. Indeed, for each instance of the attributes, i.e., each pair of [interarrival time - packet size], we compute the mean and standard deviation on a predetermined sequence of lines, for both the interarrival time and the packet size. The new data set of features is then provided to the classifier.\\\\
Overall, PiNcH consists of four (4) phases, that are: (i) \emph{WiFi Radio Eavesdropping}; (ii) \emph{Attributes Extraction}; (iii) \emph{Feature Generation}; and (iv) \emph{Classification}. The operations executed in each of them are described below. \\\\
In the \emph{WiFi Radio Eavesdropping Phase}, we collect a large number of measurements, using standard WiFi eavesdropping equipment. Specifically, we first collected a dataset constituted by samples from a specific traffic profile that we want to detect. We denote with $N$ the number of packets included in such a dataset \\\\ 
{\bf Attribute Extraction.} During the \emph{Attribute Extraction Phase}, for each n-th packet, we denote with $a_n$ the arrival time of the packet, i.e., the time when the packet was received, and with $p_n$ the corresponding packet size. Also, we denote the source MAC address of the packet as $s_n$ and the destination MAC address as $d_n$. Thus, each received packet is represented by a vector in the form $\left[s_n, d_n, a_n, p_n \right]$. We denote this data-set as $A$, being characterized by a label value $y_n = 0$.\\
For training purposes, we also collect an additional trace, containing wireless traffic related to another profile, different from the one considered in the data-set $A$. 
Each packet in this trace is represented by the attributes described before, and thus it is a vector in the form $\left[s_j, d_j, a_j, p_j \right]$.  We assume a total number of $J$ packets, and we denote this data-set as $B$, being characterized by a label value $y_j = 1$.\\
Note that this methodology can be extended to include further profiles of traffic of interest, for a total number of labels equal to $L$, where each profile can be distinguished from the others for the specific value of the associated label. For ease of discussion and without loss of generality, in the following we assume two classes, and thus a binary problem ($L=2$), with $y=\left[ 0, 1 \right]$.\\\\
{\bf Feature Generation.} In the \emph{Feature Generation} phase, for all the data-sets, we extracted the interarrival times of the packets. Considering the generic n-th packet of the data-set $A$, we compute the interarrival time between the n-th packet and the $\left( n+1 \right)$-th packet, as $t_n = a_{n+1} - a_n$, and we replace the arrival time $a_n$ with the interarrival time $t_n$. As a result, the trace $A$ contains $N-1$ vectors in the form $\left[s_n, d_n, t_n, p_n \right]$, while the trace $B$ contains $J-1$ vectors in the form $\left[s_j, d_j, t_j, p_j \right]$. \\
Starting from these traces, we create new data-sets, by extracting a number of features $F=6$ for each packet, where each f-th feature ($f=1, \dots, 6$) is computed as described below. With reference to the trace $A$, the following steps are executed:
\begin{itemize}
    \item The first feature ($x_1$) is the packet size, namely $x_{n,1} = p_n$;
    \item the second feature ($x_2$) is the interarrival time between the n-th packet and the $\left( n+1 \right)$-th packet, namely $x_{n,2} = t_{n+1} - t_n$;
    \item the third feature ($x_3$) is the mean packet size, computed over a vector having size $W$, namely $x_{n,3} = \frac{1}{W} \cdot \sum_{n=1}^W p_n$;
    \item the fourth feature ($x_4$) is the mean interarrival time, computed over a vector having size $W$, namely $x_{n,4} = \frac{1}{W} \cdot \sum_{n=1}^W t_n$;
    \item the fifth feature ($x_5$) is the standard deviation of the packet size, computed over a vector having size $W$, namely $x_{n,5} = \sqrt{ \frac{ \sum_{n=1}^W \left( p_n - x_{n,3} \right) }{W} }$;
    \item the sixth feature ($x_6$) is the standard deviation of the interarrival time, computed over a vector having size $W$, namely $x_{n,6} = \sqrt{ \frac{ \sum_{n=1}^W \left( t_n - x_{n,4} \right) }{W} }$;
    \item finally, for each n-th packet, we insert a label ($y_n$), where $y_n = \left[ 0, 1 \right]$, indicating the specific data-set the packet is related to.
\end{itemize}
Note that the above process is repeated also for the trace $B$. The result of the \emph{Feature Generation} phase is the creation of two matrices, namely $D_A$ and $D_B$, containing the features and the labels of all the packets in both the data-sets.
We mix the matrices $D_A$ and $D_B$, creating a single matrix $D$, having a total number of $M=N+J$ packets, in the form depicted in the following Eq.~\ref{eq:matrices}. \\
\begin{gather}
    \label{eq:matrices}
    \mathbf{D} = 
    \left\{
    \begin{tabular}{ccccc}
    $x_{1,1}$ & $x_{1,2}$ & \dots & $x_{1,6}$ & $y_{1}$ \\
    $x_{2,1}$ & $x_{2,2}$ & \dots & $x_{2,6}$ & $y_{2}$ \\
    $\vdots$& $\vdots$& $\ddots$&$\vdots$ \\
    $x_{m,1}$ & $x_{m,2}$ & \dots & $x_{m,6}$ & $y_{m}$ \\
    $\vdots$& $\vdots$& $\ddots$&$\vdots$ \\
    $x_{M,1}$ & $x_{M,2}$ & \dots & $x_{M,6}$ & $y_{M}$ \\
    \end{tabular} 
    \right\} = \left\{ \mathbf{x}, \mathbf{y} \right\}
\end{gather}
This matrix is given in input to the \emph{Classification Phase}, and it is used to classify and predict the class of the packets. \\\\
{\bf Classification.} The goal of the \emph{Classification Phase} is to build a classifier, which predicts the label $y_n$ from the features $x_n$, based on the data-set $D$, given an ensemble of classifiers $h = \left\{ h_1(\mathbf{x}), h_2 (\mathbf{x}), \dots, h_K (\mathbf{x}) \right\}$, where the classifiers $h$ are decision trees, and therefore, the ensemble is a \emph{Random Forest}.\\
Specifically, a decision tree is a classification tool leveraging a tree-like graph or decision model, including event probability, resource costs, and utility. It is a useful tool to represent an algorithm containing only conditional control statements. Formally, a classification tree is a decision tree where each node has a binary possible decision, depending on whether the input feature $x_f$ is subject or not to the condition $x_f < a$, being $a$ a threshold parameter of the decision tree. The top node of the decision tree is defined as the root node, and it contains the whole data-sample. Then, the data-sample is binarily sub-divided into smaller parts, namely sub-samples, where each sub-sample satisfies (or not) the condition defined by the threshold. The criterion is that the subdivision continues until each sub-group has only a single label, thus being related to a single class, or there is not any further sub-division that improves the actual situation.\\
There are several algorithms for constructing decision trees, i.e., to decide the value of the threshold $a$ on each node of the tree. These techniques work top-down, by choosing a threshold at each step that best splits the set of items. Some examples include Gain Impurity Maximization, Information Gain Maximization, and Variance Reduction. In our approach, we selected the Information Gain Maximization approach, whose main logic is described in the following.
\begin{itemize}
    \item We first compute the entropy of the data-set, namely $H(T)$, as in the following Eq.~\ref{eq:entropy}.
    \begin{equation}
    \label{eq:entropy}
        H ( T ) = - \sum_{l=0}^{L-1} P_l \cdot \log_2 P_l,
    \end{equation}
    where $P_l$ is the probability of each class present in the child node that results from a split in the tree, with $\sum_{l=0}^{L-1} P_l = 1$~\cite{frank2011}.
    \item Then, given a threshold value $a$, the Information Gain $IG(T,a)$ derived from the split of trees with the threshold $a$ is defined as in the following Eq.~\ref{eq:ig}.
    \begin{flalign} \nonumber
        \label{eq:ig}
        IG (T,a) &= H(T) - H(T|a) = &&\\
        &= - \sum_{l=0}^{L-1} P_l \cdot \log_2 P_l - P(a) \cdot \sum_{l=0}^{L-1} P(l|a) \cdot \log_2 P(l|a).
    \end{flalign}
    \item The specific threshold value $a_t$, selected at the node $t$ of the tree, is the one that maximizes the information gain in the dataset $T$, as in Eq.~\ref{eq:max_ig}.
    \begin{equation}
        \label{eq:max_ig}
        a_t = \max_{a \in vals(x_k)} IG (T,a),
    \end{equation}
    where the notation $vals(x_k)$ refers to the set of possible values for the feature $x_k$.
\end{itemize}
The above strategies are used to obtain the set of decision trees that best fit the data, i.e., by maximizing the information gain. Note that the specific parameters of the \emph{best} decision trees are not fixed, but they can vary due to the randomness in the usage order of the features and according to the specific strategy used to obtain the threshold at each node of the tree. More details on further techniques and optimizations used in the implementation of decision trees can be found in~\cite{frank2011} and~\cite{Rokach2008}.\\
Now, considering a large number of decision trees, a \emph{Random Forest} is a generalized classifier that considers many decision trees together. Formally, a Random Forest is a classifier based on a family of classification trees $h = \left\{ h( \mathbf{x}| \Theta_1), \dots, h( \mathbf{x}| \Theta_K) \right\} $, where $\Theta_k$ are the parameters of the classification trees, that are randomly chosen from a model random vector $\Theta$. Note that these parameters refer to the variables of the decision tree, including the structure of the tree, the number of layers, and the configuration of the threshold values. \\
Assuming the final classification of $\mathbf{x}$ is denoted by $f(\mathbf{x}) = \mathbf{y}$, and each decision tree $h_k (\mathbf{x})$ casts a vote for the most popular class that is $y_k (\mathbf{x})$, we have that $f(\mathbf{x})=\mathbf{y}$ is the most popular classification of $x$ in the ensemble of decision trees, i.e., formally:
\begin{equation}
    \label{eq:rf}
    f(\mathbf{x}) = \mathbf{y} = \max_{l_0, l_1} \left\{ \left[ \sum_{k=1}^K \left( h_k(\mathbf{x}) == 0 \right) , \sum_{k=1}^K \left( h_k(\mathbf{x}) == 1 \right)
    \right] \right\}.
\end{equation}
We refer the interested readers to the technical works on Decision Trees and Random Forest classification by the authors in~\cite{breiman2001} and \cite{Paul2018} for details about the classification accuracy and further improvements on specific estimation techniques. \\
Finally, we highlight that, for the Random Forest classification algorithm, we used the 10-folds cross-validation method, i.e., a technique to evaluate the classifier performance by partitioning the original sample into a training set (9 randomly chosen folds) to train the model, and a test set to evaluate it (the remaining fold from excluding the 9 previously chosen). This is a standard technique used in data mining processes, and further details can be found in~\cite{rodriguez2010}.}

%
%

\section{Drone scenario identification: Is it flying?}
\label{sec:scenario_identification}
\begin{figure}[htbp]
\includegraphics[width=.75\columnwidth]{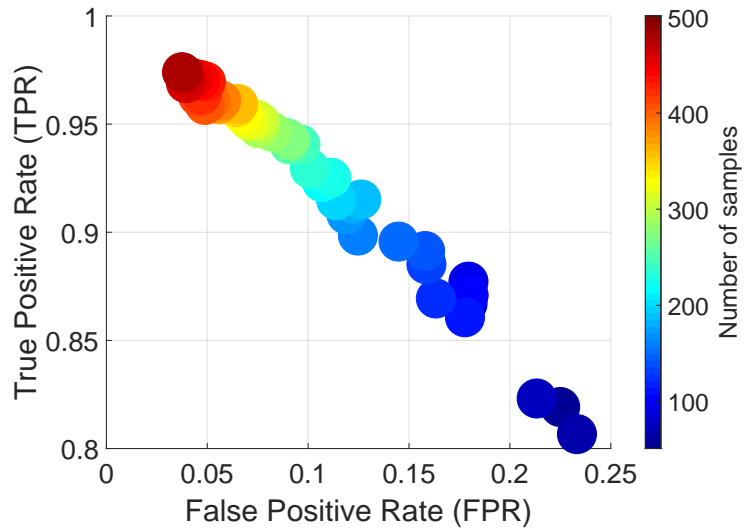}
\centering
\caption{\ac{TPR} and \ac{FPR} to detect the state of the drone as a function of the number of eavesdropped packets.}
\label{fig:mov_steady_ROC}
\end{figure}

In this section, we introduce the methodology used by \sol\ to identify the current state of the drone, i.e., Steady or Movement. To this end, we structured the detection system over the following steps: eavesdropping the WiFi spectrum, collecting packets, generating the associated features, and classifying the incoming traffic, as discussed in Section \ref{sec:classification_methodology}. For each instance in the features set, we challenge the classifier to identify the state of the drone. 
In particular, we consider 5 different metrics: \ac{TP}, \ac{FP}, \ac{FN}, \ac{TN}, and the overall \ac{SR} as the number of correct classifications (TP+TN) divided by the total number of classifications (TP+TN+FP+FN). Table \ref{table:static_mov} depicts the results of the 10-fold cross-validation, assuming the three aforementioned classifiers and metrics with the features computed over a reference time window of 200 consecutive samples. Link fingerprinting achieves the worst performance, while the unique flows, i.e., RC to Drone and Drone to RC, behave almost in the same way. Considering the one-way links, we observe that both FP and FN are less than 9\%. Finally, for each configuration, we report (in percentage) the overall \ac{SR} being equal to the total number of correctly classified instances.
\begin{table}[htbp]
\centering
\caption{Detection of drone's state considering different flows. All the values are in percentages [\%], while the Random Forest (RF) classifier is used as the classification tool.}
\label{table:static_mov}
\begin{tabular}{|l|l|c|c|c|c|c|}
\hline
{\bf Classifier} & {\bf Flow} & {\bf TP} & {\bf FP} & {\bf FN} & {\bf TN} & {\bf SR} \\
\hline
\hline
\multirow{3}{*}{RF} & RC to Drone & 92.37 & 7.63 & 8.65 & 91.35 & 91.86\\
                    & Drone to RC & 92.98 & 7.02 & 6.89 & 93.11 & 93.05\\
                    & Link  & 88.15 & 11.85 & 8.57 & 91.43 & 89.69\\
\hline
\end{tabular}
\end{table}
\\
{\bf Detection delay.} We now consider the time to detect the state of the drone. Given the results of the previous section, we assume the Random Forest classifier using the 10-folds cross-validation method and the unicast traffic collected from the two links, i.e., RC to Drone and Drone to RC. Figure \ref{fig:mov_steady_ROC} shows the \ac{ROC} curve associated with the aforementioned configuration while varying the number of samples used for the generation of the features. Indeed, as in the previous case, we considered the inter-arrival time, the packet size, and their mean and standard deviation computed over partially overlapping sliding windows of size spanning between 50 and 500 samples.

Increasing the number of samples used to compute the mean and the standard deviation significantly improves the performance of the detection process. To provide few reference performance indicators, considering 200 samples, we can achieve 0.91 of \acl{TPR} and 0.11 of \acl{FPR}, while increasing the number of samples to 400 leads to 0.96 of TPR and 0.05 of FPR. 

The number of samples used to compute the mean and the standard deviation for each of the above traffic features is proportional to the detection delay, i.e., the time window required to collect such a number of samples.

\begin{figure}[htbp]
\includegraphics[width=.75\columnwidth]{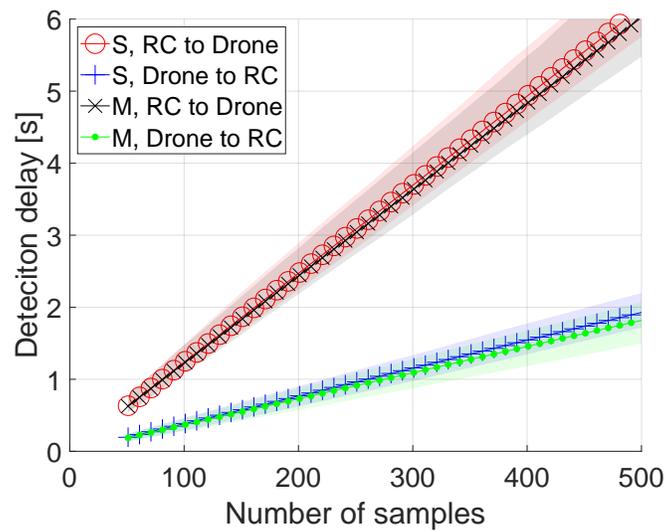}
\centering
\caption{Relationship between the number of samples and the detection delay, for each of the four flows.}
\label{fig:detection_delay_mov}
\end{figure}

Figure \ref{fig:detection_delay_mov} shows the quantile 0.05, 0.5, and 0.95 associated to the detection delay for each of the four flows, as a function of the number of samples. It is possible to notice that the flows from the RC to the drone (and similarly, from the drone to the RC) exhibit the same performance, independently from the particular operational mode of the drone, i.e., being it steady or moving. For the Drone to RC flow, being characterized by the highest throughput, the detection delay is smaller than the RC to drone flow. 

Given the linear relationship between the number of samples and the detection delay, a good trade-off between detection performance and delay can be estimated in 200 samples (FPR=0.11, TPR=0.91) being equivalent to about 2.45 seconds and 0.73 seconds of eavesdropping time, for the \emph{RC to drone} and the \emph{drone to RC} flow, respectively. Better performance can be achieved using 400 samples (FPR=0.05, TPR=0.96) while incurring in a detection delay of about 4.8 seconds for the RC to drone flow and 1.5 seconds for the drone to RC flow.

\begin{figure}[htbp]
    \includegraphics[width=\columnwidth]{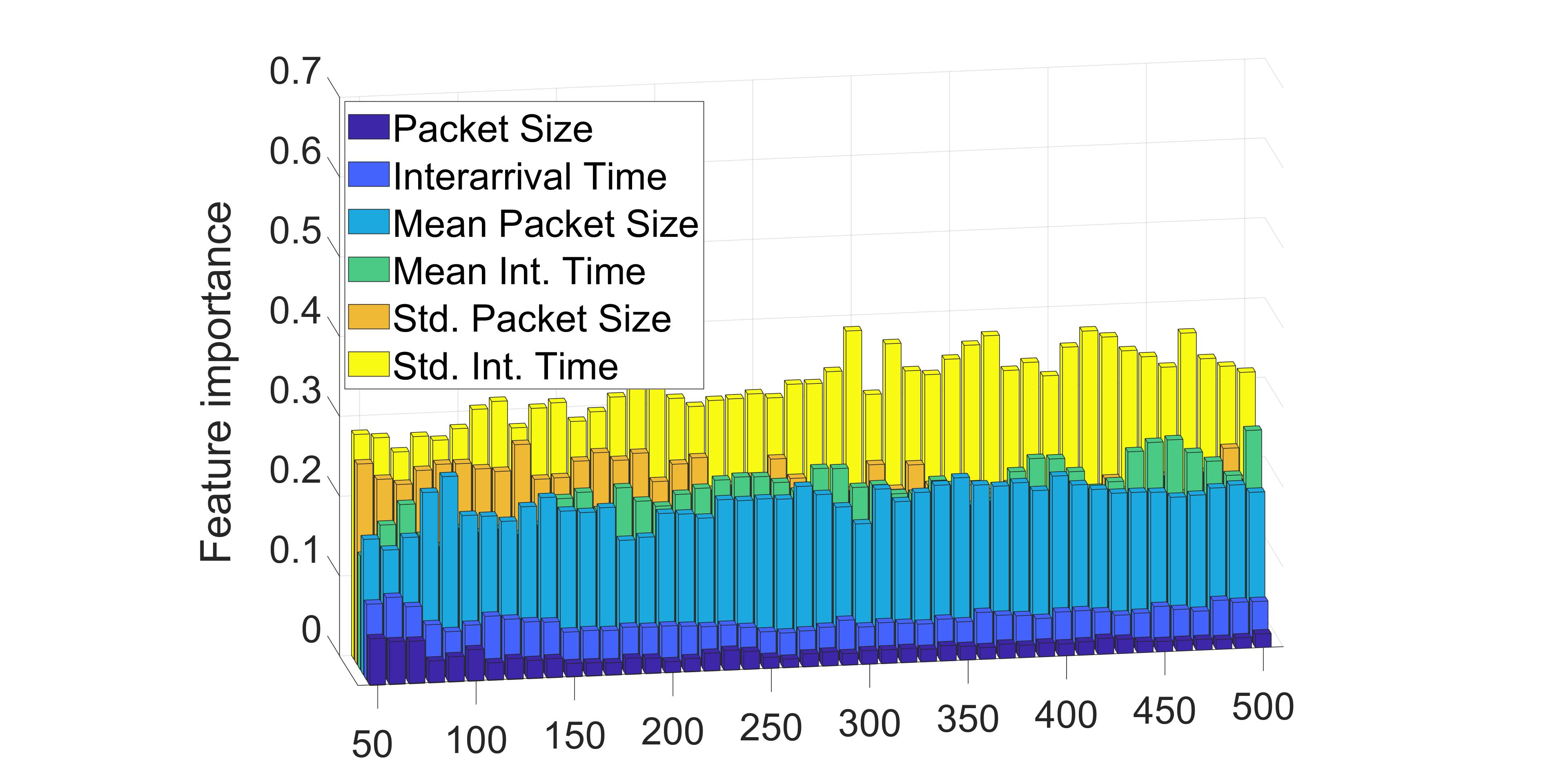}
    \centering
    \caption{Normalized Feature Importance with different sample size (and detection delay).}
    \label{fig:feat_imp_mov}
\end{figure}

To provide further insights, we also investigated the relative impact, i.e., the weight, that each feature has in the model. Figure \ref{fig:feat_imp_mov} summarizes our analysis, showing the normalized \emph{feature importance} in the above scenario, as it has been obtained via the Machine Learning Toolbox of MATLAB R2019a.

The standard deviation of the interarrival time between the packets (yellow bars) has the highest impact, showing a feature importance value that is almost 35\%. Among the other features, we report that the mean of the interarrival time, as well as the ``raw'' interarrival time between packets, always emerge as the second important feature. 

These considerations will be further leveraged later on in Section \ref{sec:robustness}, where we will discuss the effect of evasion strategies that an attacker can deploy.

\section{Detecting a drone into the wild}
\label{sec:detecting_a_drone}
This section introduces the techniques and methodologies used by \sol\ to detect the presence of a drone in different scenarios. We consider five different traces from the CRAWDAD data-set \cite{crawdad} and one trace collected from a crowded outdoor area in Doha, Qatar. We considered these traces as our reference scenarios for testing the presence of the drone. The traces have been selected to guarantee the maximum scenario heterogeneity and the presence of different WiFi network patterns.
\begin{table}[htbp]
\centering
\caption{WiFi traces description.}
\label{table:wifi_traces}
\begin{tabular}{|P{0.02\columnwidth}|l|c|c|}
\hline
 {\bf ID} & {\bf Description} & {\bf Size (B)} & {\bf Ref.} \\
 \hline
 S1 &  \makecell[l]{Wireless LAN traffic trace collected from \\ PSU (Portland State University) Library.} & 89905 & \cite{t1} \\
  \hline
 S2 & \makecell[l]{Wireless LAN traffic trace collected from \\ PSU (Portland State University) Cafeteria.} & 131301 & \cite{t1}  \\
  \hline
 S3 & \makecell[l]{Wireless LAN traffic trace collected from \\ a large outdoor area in downtown Portland.} & 29795 &\cite{t1}  \\
  \hline
 S4 & \makecell[l]{Tcpdump trace from the wireless network \\ at a three-day computer-science conference.}  & 110492 &\cite{t4} \\
  \hline
 S5 & \makecell[l]{Wireless probe requests collected at a \\ political meeting in Rome, Italy.} & 11799 & \cite{t5} \\
 \hline
 S6 & \makecell[l]{Measurement from an outdoor area in \\ Doha, Qatar.} & 82651 &  \\
 \hline
\end{tabular}
\end{table}
Table \ref{table:wifi_traces} shows the traces we selected to test the presence of the drone, together with a brief description of them.

We carefully analyzed the content of the traces and we selected a time period of 10 minutes from each trace, to provide a balanced data set, where the duration of the traces is the same. We consider several scenarios: a library (S1), a cafeteria (S2), outdoor areas (S3, S6), a computer science conference (S4), and a political meeting (S5). We selected both indoor (S1, S2, S4, S5) and outdoor scenarios (S3, S6), with the presence of both smart-phones and laptops, and characterized by different moving patterns. 
\\
Given the performance of the previous configurations, we select the Random Forest classifier, the Drone to RC flow and the 10-fold cross-validation method. The Drone to RC flow guarantees also a more realistic scenario, being usually the drone more exposed to the eavesdropping equipment compared to the remote controller.
Moreover, in this section we consider the features computed over a time window of 20 samples, being equivalent to less than 0.28 seconds of channel eavesdropping (quantile 0.95 value). This leads to a total number of 2,142 instances, where for each test the 214 samples were randomly chosen to be the test set (10\%), while the remaining samples were used for training.

In addition, we only consider the scenario related to the moving drone, being more suitable for the detection problem introduced by Figure \ref{fig:model}. Finally, we mixed each of the above traces with a standalone trace of the moving drone.

\begin{figure}[htbp]
\includegraphics[width=.9\columnwidth]{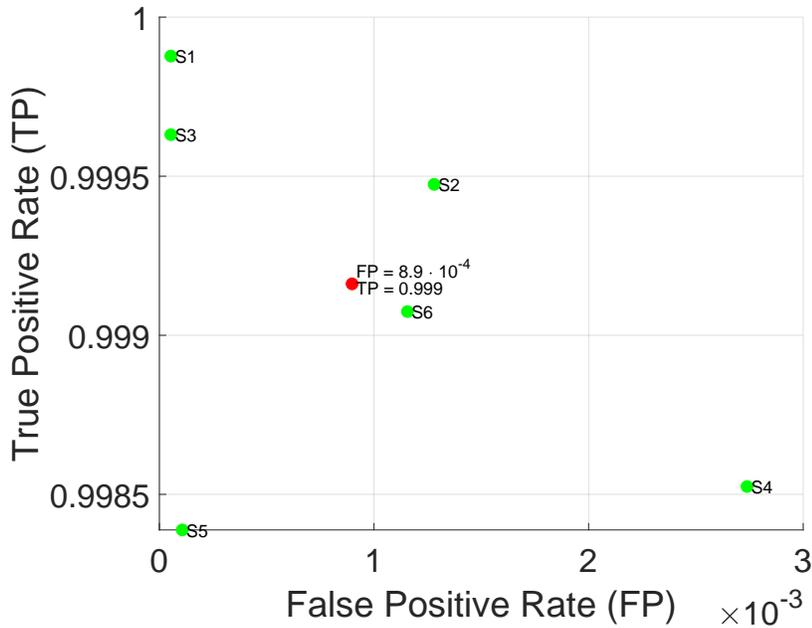}
\centering
\caption{TP and \ac{FP} rates for the detection of the drone in the 6 scenarios, i.e., S1, S2, S3, S4, S5, and S6. The red circle represents an average estimation of all the scenarios.}
\label{fig:yn_ROC}
\end{figure}
The green circles in Figure \ref{fig:yn_ROC} show the \ac{TPR} as a function of the \ac{FPR} for each of the previously introduced scenarios S1, S2, S3, S4, S5, and S6. We highlight that the Random Forest classifier performs extremely well, being able to detect the presence of the drone with FP $< 0.0027$ (worst case, S4) and TP $> 0.9984$ (worst case, S5). Finally, we report an estimation of the average behaviour by the red circle characterized by FP = $8.9 \cdot 10^{-4}$ and FN = 1-TP = $0.9 \cdot 10^{-4}$.
%
%

We further analyze the detection delay for the previously introduced scenarios S1, S2, S3, S4, S5, and S6. The detection delay is a particularly relevant metric, since it provides an estimation of the time required to detect the presence of the drone.

Recalling the scenario introduced in Figure \ref{fig:model}, we observe that the detection delay significantly affects the probes layout and the surveillance area. Indeed, large detection delays do imply larger reaction time and, in turn, a much larger surveillance area to guarantee enough reaction time. Conversely, short detection delays allow for faster reaction time and shorter distances between the target and probes.

\begin{figure}[htbp]
\includegraphics[width=.79\columnwidth]{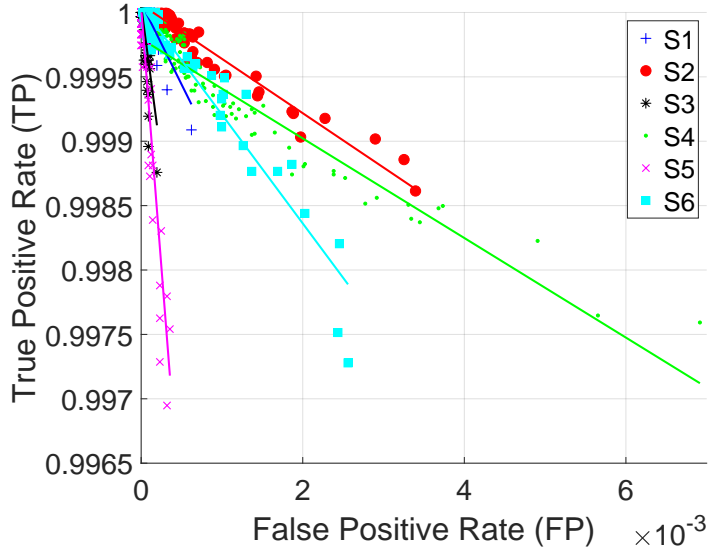}
\centering
\caption{\ac{TPR} and \ac{FPR} for the 6 scenarios while varying the number of eavesdropped samples. Solid lines represent the linear regression for each of the considered scenario.}
\label{fig:detection_delay}
\end{figure}

We adopted the following configuration: Random Forest classifier, features computed over consecutive samples spanning between 5 and 200, 10-fold cross-validation method, and finally the scenarios S1, S2, S3, S4, S5, and S6. Figure \ref{fig:detection_delay} shows the \ac{TPR} as a function of the \ac{FPR}, varying the number of samples for the 6 different scenarios. We observe that the Random Forest classifier is effective in detecting the presence of the drone in all the scenarios, i.e., for the majority of the cases FP $< 3 \cdot 10^{-3}$ and TP $> 0.997$, even for the worst-case scenario where we consider a detection delay based on only five samples. We highlighted the trends by computing the linear regression for each scenario, i.e., solid lines in Figure \ref{fig:detection_delay}.

\begin{figure}[htbp]
\includegraphics[width=.8\columnwidth]{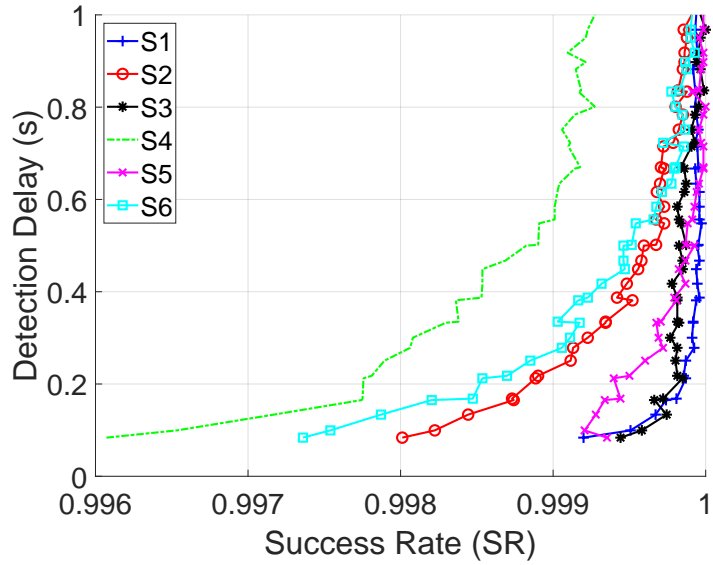}
\centering
\caption{Detection delay as a function of the overall \ac{SR} for the 6 different scenarios.}
\label{fig:detection_delay_SR}
\end{figure}

To estimate the detection delay, we vary the number of samples required to generate one instance of the features and we compute the detection delay as a function of the overall \ac{SR} defined as SR = (TP+TN) / (TP+TN+FP+FN) as depicted in Figure \ref{fig:detection_delay_SR}. 
The detection delay is directly proportional to the number of samples considered for the computation of the standard deviation and the mean of the packet size and the inter-arrival time. Such a delay spans between 0, i.e., only one packet is considered to infer on the presence of the drone, and 1 second. 
Moreover, the detection delay might significantly increase when an overall \ac{SR} greater than 0.999 is required. Nevertheless, we observe that the Random Forest classifier guarantees a detection delay less or equal to half of a second, assuming a SR less than 0.999 for all the scenarios. The scenarios slightly affect the performance of the classifier: S4 (Computer Science Conference) is the worst, S2 and S6 (Cafeteria and Outdoor Doha) behave very similarly, and lastly, S1, S3, and S5 (Library, Outdoor Portland, and Political Meeting) have the best performance. 

\section{Drone movements identification}
\label{sec:movement_identification}
In this section, we consider the problem of identifying the specific movements performed by the drone. We consider 7 different movements as depicted in Table \ref{table:movements}. To collect the network patterns associated with each drone movement, we unscrewed the drone's propellers and we collected about 9400 packets (1 minute) for each drone movement. We used the Random Forest classifier, and the 10-folds classification method. Differently from our previous analysis, in this scenario, we do not use the Drone to RC flow but we consider the overall link, since preliminary experiments indicated that better performance could be seized.
\begin{table}[h]
\centering
    \caption{Remote controller commands and corresponding drone movements.}
    \centering
    \label{table:movements}
    \begin{tabular}{|l|l|}
        \hline
        {\bf Stick position} & {\bf Description} \\
        \hline
        Pitch down &  \makecell[l]{The right stick is pushed forward and the drone \\ moves forward.} \\
        \hline
        Pitch up &  \makecell[l]{The right stick is pushed backward and the drone \\ moves backward.} \\
        \hline
        Roll left &  \makecell[l]{The right stick is pushed left and the drone \\ moves left.} \\
        \hline
        Roll right &  \makecell[l]{The right stick is pushed right and the drone \\ moves right.} \\
        \hline
        Throttle up & \makecell[l]{The left stick is pushed forward and the drone \\ increases its altitude.} \\
        \hline
        Yaw left &  \makecell[l]{The left stick is pushed left and the drone \\ rotates left.} \\
        \hline
        Yaw right &  \makecell[l]{The left stick is pushed right and the drone \\ rotates right.} \\
        \hline
    \end{tabular}
\end{table}

\begin{figure}[htbp]
    \includegraphics[width=.8\columnwidth]{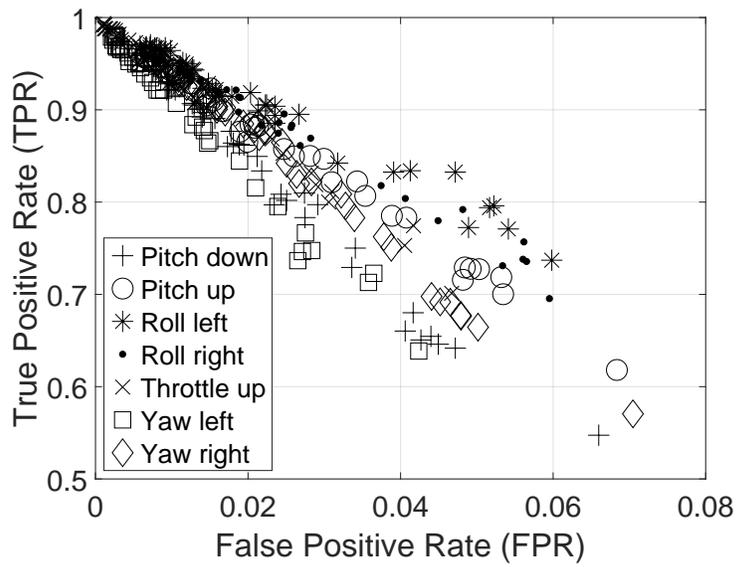}
    \centering
    \caption{\ac{TPR} and \ac{FPR} associated to each movement of the drone.}
    \label{fig:mov_identification}
\end{figure}

Figure \ref{fig:mov_identification} shows the \ac{TPR} as a function of the \ac{FPR} associated to the seven aforementioned movements. For each movement, we consider different sample sizes to generate the features spanning between 50 and 500 consecutive packets. We observe that movement identification is significantly affected by the sample size; indeed, TPR and FPR span between [0.001, 1] and [0.07, 0.56], respectively. 

\begin{figure}[htbp]
    \includegraphics[width=.8\columnwidth]{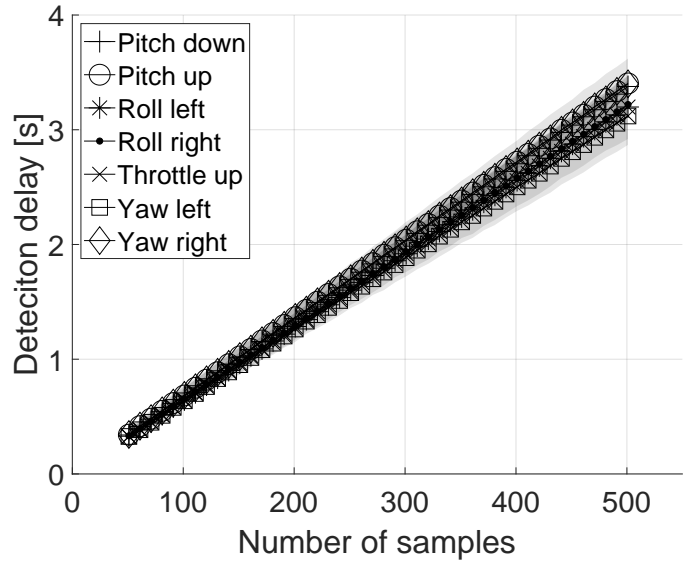}
    \centering
    \caption{Detection delay associated to each movement of the drone.}
    \label{fig:mov_identification_3}
\end{figure}

We also investigate the detection delay related with each sample size. The results are provided in Figure \ref{fig:mov_identification_3}.

It is possible to note that in all the cases the quantile 0.05, 0.5, and 0.95 of the detection delays are almost overlapping, demonstrating that the interarrival time between packets shows a similar profile over the time, independently of the particular movement.  

\begin{figure}[htbp]
    \includegraphics[width=.8\columnwidth]{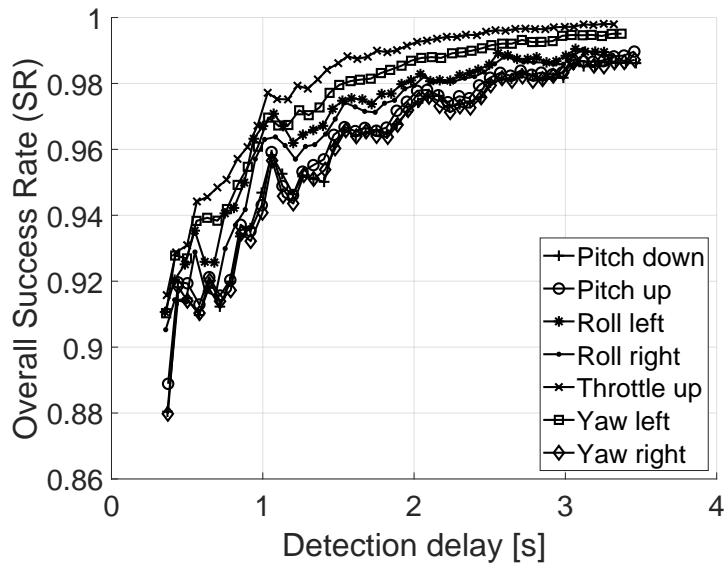}
    \centering
    \caption{Overall \ac{SR} of the Random Forest classifier as a function of the detection delay.}
    \label{fig:mov_identification_4}
\end{figure}

Finally, Figure \ref{fig:mov_identification_4} shows the overall \acfi{SR} as a function of the detection delay. 

It is worth noting that a total number of just 220 samples, i.e., approximately 1.5 seconds, are necessary for the Random Forest classifier to guarantee a value of the SR $> 0.95$ in the discrimination of any movement performed by the drone. Thus, it is enough to either push or pull the stick of the remote controller for a time frame longer than 1.5 seconds to allow \sol\ to discriminate the specific movement performed by the drone.

\section{Assessing the Robustness of \sol}
\label{sec:robustness}

In this section we discuss the robustness of the proposed detection strategy against packet loss and evasion strategies, being these the dominant factor that could affect the effective deployment of \sol.

{\bf Drone Detection and Packet Loss}. In Section \ref{sec:detecting_a_drone} we showed the remarkable performance of \sol\ in detecting the presence of a drone in several scenarios, being them indoor or outdoor.

Since our detection scheme is (partially) based on the interarrival time between packets, and such a feature is the most dominant, packet loss phenomena can have an impact on the detection performance, especially at long distances. In fact, the intuition suggests that when the drone-RC communication link becomes long enough, packets can be lost, leading to larger interarrival times and decreased detection rates.

\begin{figure}[htbp]
    \includegraphics[width=\columnwidth]{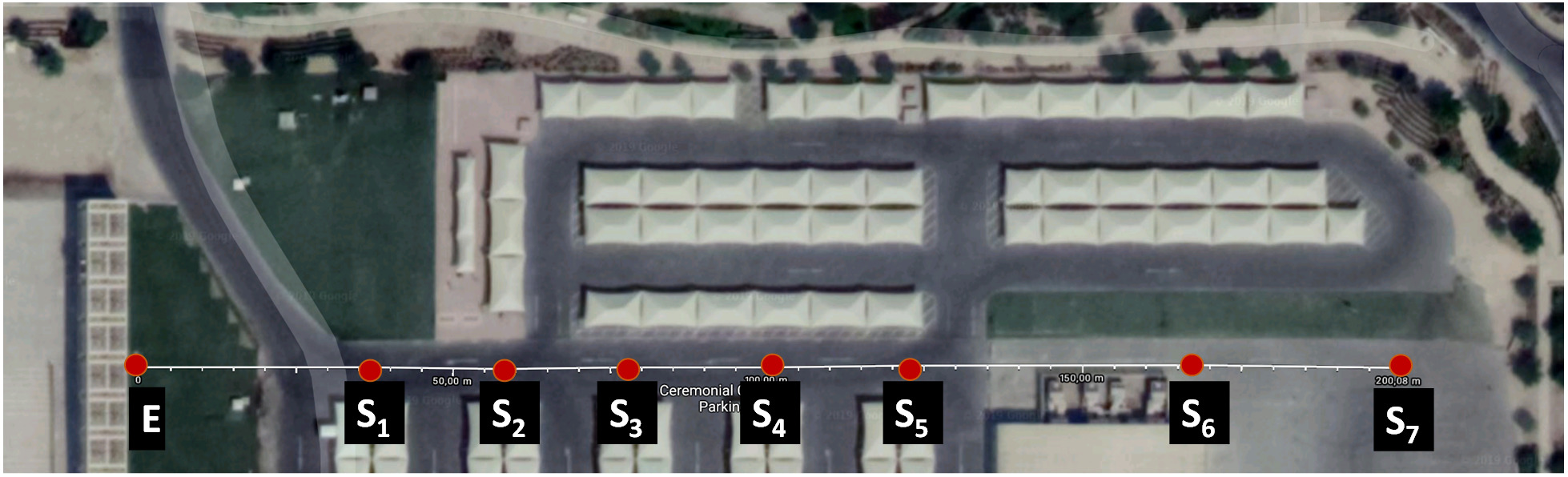}
    \centering
    \caption{Map of the location where outdoor experiments have been performed (taken from Google Earth).}
    \label{fig:map}
\end{figure}

To provide further insights on this phenomenon, we investigated the detection rate of \sol\ in outdoor scenarios, by placing the RC-drone communication link at increasing distances from the location of the eavesdropping equipment. Figure \ref{fig:map} shows the deployment of our tests, where \textit{E} is the location of the WiFi probe, while the locations \textit{$S_1$, $S_2$, $S_3$, $S_4$, $S_5$, $S_6$,} and \textit{$S_7$} refer to the tested locations of the RC-drone communication link (i.e, distances of 30, 50, 70, 95, 115, 170, and 200 meters from the drone, respectively). 

We notice that the location is full of obstacles, providing realistic shadowing effects affecting outdoor application scenarios.

We first trained \sol\ on the profile of the traffic acquired at a distance equal to $0.2$ m, and then we tested it on the acquisitions at the various distances, by considering the whole traffic acquired on the communication link mixed with the various traces discussed in Sec. \ref{sec:detecting_a_drone}. Average results are summarized in Tab. \ref{tab:p_loss}.
\begin{table*}[htbp]
\caption{Packet Loss and Detection Rate at increasing distances from the position of the eavesdropping equipment.}
\centering
\label{tab:p_loss}
\begin{tabular}{|P{1.7cm}|P{1.7cm}|P{1.7cm}|P{1.9cm}|}
\hline
 {\bf Location } & {\bf Distance [m]} & {\bf Packet Loss [\%]} & {\bf Detection Rate [\%]} \\
 \hline
 $S_1$ & 30 & 0 & 99.999 \\
 \hline
 $S_2$ & 50 & 21.1 & 99.69 \\
 \hline
 $S_3$ & 70 & 33.2 & 99.23 \\
 \hline
 $S_4$ & 95 & 13.3 & 99.68 \\
 \hline
 $S_5$ & 115 & 16.7 & 99.18 \\
 \hline
 $S_6$ & 170 & 74.8 & 97.43\\
 \hline
 $S_7$ & 200 & 73.8 & 99.68\\
 \hline
 \end{tabular}
\end{table*}

We highlight that, despite the increasing packet loss percentage at increasing distances, \sol\ is still able to identify the presence of the drone with outstanding accuracy ($\ge 97\%$), being robust to packet loss up to 74.8\%. These results suggest that the detection range could be further extended by using specialized equipment, such as directive antennas, to provide effective detection even to larger distances.

{\bf Evasion Strategies}. The previous sections highlighted the remarkable performance achieved by \sol\ for identifying a drone. By resorting to the acquisition of packet size and interarrival times via general-purpose eavesdropping equipment, \sol\ can identify the presence of a drone in several scenarios with outstanding accuracy, requiring just a negligible detection delay.

In this section, we assume the adversary is aware of the deployment of \sol, and therefore, she implements a strategy to escape the detection, by modifying on purpose the profile of the features exploited to detect the presence of the drone. 

For instance, being aware that the interarrival time is the most important feature of our drone detection solution, the attacker can delay the delivery time of the packets, in order for the eavesdropping equipment to record a profile of interarrival times that is different from the expected one, possibly leading to incorrect classification. 

\begin{figure}[htbp]
    \includegraphics[width=.79\columnwidth]{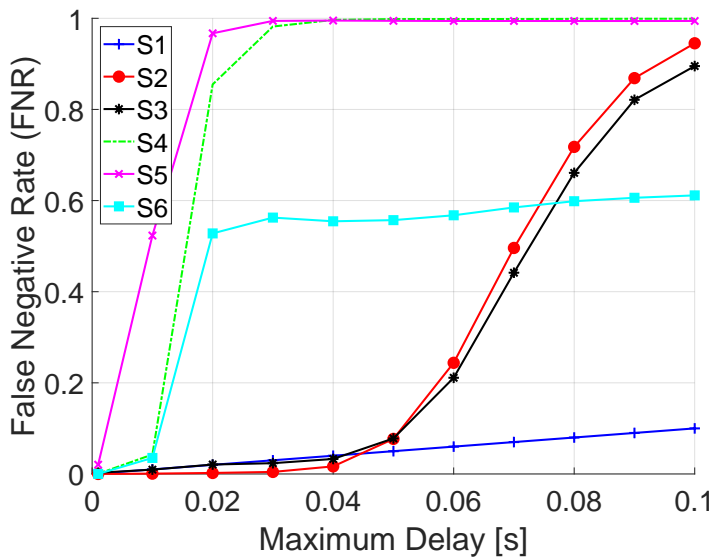}
    \centering
    \caption{\acl{FNR} as a function of the maximum delay $\Delta$ in the communication pattern.}
    \label{fig:evasion}
\end{figure}

We emulated the evasion attack performed by the attacker, by summing random delays extracted from a uniform distribution $[0, \Delta]$ to the interarrival times, with $\Delta$ being arbitrarily large, up to $0.1$ s. By focusing on the six scenarios tackled in Section \ref{sec:detecting_a_drone} and assuming a window size of 21 samples, Figure \ref{fig:evasion} shows the \ac{FNR}, i.e., the number of samples incorrectly classified as ``no-drone'', although being from the drone communication channel, while increasing the maximum delay $\Delta$.

We observe that the effectiveness of the evasion attack strictly depends on the particular scenario, i.e., on the features of the surrounding encrypted traffic.

With reference to the scenarios $S5$ and $S4$, we notice that the attacker could delay the delivery of the packets (either on the RC or on the drone) of a maximum value of $20$ ms to effectively escape the detection ($FNR \geq 0.8$ when the maximum injected delay is up to 0.2 seconds). This finding is due to the ``distance'' between the distribution of the traffic of the RC-drone communication link and the distribution of the surrounding traffic, which are close to each other.

Regarding the scenarios $S2$ and $S3$, a greater value of the maximum delay is necessary to effectively avoid the detection. Specifically, we notice that the attacker has to delay packets with a maximum delay of $60$ ms to cause the incorrect classification of about $25\%$ of the samples, while larger values of the FNR are obtained with a maximum delay of 80 ms, where FNR reaches values of about 70\%.

At the same time, evasion attacks seem not to be effective in the case of the scenario $S1$, where delaying the packets of 0.1 s lead to FNR values less than $15\%$. 

Finally, in scenario $S6$, FNR reaches the value of 53\% for a maximum delay of 20 ms, and then, it remains almost constant for increasing delays. In this case, the profile of the ``modified'' traffic is equally different from the two known distributions, leading to an almost random decision.

Overall, the results reported above should be assessed also considering the maneuverability, latency, and response time of the drone itself. While very short delays of the packets do not affect the communication link and the drone maneuverability, large values (e.g., 100 ms) reduce the response time and the capability of controlling the drone, especially when it is moving at high speeds and characterized by a heavy weight (high inertia)--- this is common for drones carrying payloads. Note also that the on purpose introduced delay values sum up to the intrinsic delay of the RC-drone communication link, and therefore, further increase the overall latency of the communication link. Available reports on the subject indicate that military drones can tolerate up to a maximum latency of 1 ms, while for commercial drones values of delays exceeding 60-70 ms significantly affect the maneuverability of the drone and have an impact on the human perception of the performance \cite{TNOReport}.   

Thus, depending on the specific scenario and the setup of the RC-drone communication link, the application of evasion strategies should be carefully evaluated by the attacker, trading off between the evasion of the detection and its effective capability of controlling the device, in particular at long distances.

We highlight that, despite the strategy discussed above is not the only one that can be used by the adversary to escape the detection, other strategies (such as the setup of a persistent active connection between the drone and the RC) would not guarantee any result and, in addition, they would quickly drain the battery of the drone, further reducing its operational time.

Finally, we point out that, despite the possible advantages that the adversary could gain in carefully implementing tailored evasion techniques, at this time there are no products actually implementing advanced evasion strategies.

\section{\textcolor{black}{Theoretical and Experimental Comparison}}
\label{sec:comparison}

\textcolor{black}{In this section we compare \sol\ against the solutions available in the literature at the time of this writing. Section~\ref{sec:theor_comp} provides a thorough comparison based on several system requirements, while Section~\ref{sec:perf_comp} provides an experimental comparison on real data.}

\subsection{Theoretical Comparison}
\label{sec:theor_comp}

Table \ref{tab:related} provides a comparison between our proposed approach and the closest related work in the literature, based on several system requirements.

\begin{table}[htbp]
\rotatebox{90}{
\begin{tabular}{|P{1.1cm}|P{1.3cm}|P{1.4cm}|P{1.4cm}|P{1.7cm}|P{1.7cm}|P{1.6cm}|P{1.8cm}|P{1.8cm}|P{1.8cm}|}
\hline
 {\bf Ref.} & {\bf Det. Type} & {\bf Drone OS Type} & {\bf Results Replicability} &  {\bf Drone Det. in Various Scenarios } & {\bf Status Identification} & {\bf Mov. Identification} & {\bf Det. Delay Awareness} & {\bf Packet Loss / Distance Robustness} & {\bf Study of Evasion Attacks} \\
 \hline
 \cite{Nguyen2017} & RF-based & Closed & Single Brands & \cmark & \xmark  & \xmark  & \xmark & \cmark & \xmark \\
 \hline
 \cite{fu} & RF-based  & Closed  & Single Brand & \xmark & \xmark  & \xmark  & \xmark  & \xmark & \xmark  \\
 \hline
 \cite{bisio} & Traffic Analysis & Closed & Single Brand & \xmark & \xmark & \xmark & \xmark & \xmark & \xmark \\
 \hline
 \cite{Bisio2018_TVT} & Traffic Analysis & Closed & Single Brand & \xmark & \xmark & \xmark & \cmark & \xmark & \xmark \\
 \hline
 \cite{Alipour2019} & Traffic Analysis & Closed & Single Brands & \xmark & \xmark & \xmark & \cmark & \xmark & \xmark \\
 \hline
 \cite{alipour2019_journal} & Traffic Analysis & Closed & Single Brands & \xmark & \xmark & \cmark & \cmark & \xmark & \xmark \\
 \hline
 \sol\ & Traffic Analysis & Open Source & Over 30 Brands & \cmark & \cmark & \cmark & \cmark & \cmark & \cmark \\
 \hline
\end{tabular}
}

\caption{Comparison of \sol\ with related work on passive drone detection.}
\centering
\label{tab:related}
\end{table}

On the one hand, RF-based approaches such as \cite{Nguyen2017} and \cite{fu} only accomplished the detection of a single brand of drone, whose firmware is based on a closed operating system. Thus, their results are not directly applicable to other brands. Moreover, as anticipated in Section \ref{sec:related}, these approaches usually have to rely on specific hardware, such as \acl{SDR}.

On the other hand, competing approaches based on encrypted traffic analysis are still based on closed source firmware and operating systems. In addition, none of the previous work evaluated the effectiveness of the drone detection scheme in various scenarios with real traces. Moreover, we are the first ones to prove the robustness of the detection scheme when confronting with packet loss, showing the practical applicability of encrypted traffic analysis toward the protection of critical infrastructures.

Furthermore, differently from competing approaches, in this paper, we provide an estimation of the detection performance of our methodology in the presence of evasion attacks, i.e., smart strategies where the attacker modifies the profile of the interarrival times of the packets on purpose to avoid detection. On the one hand, the effectiveness of such techniques strongly depends on the specific scenario, while on the other hand, their application could significantly decrease the response time and the maneuverability of the drone. 

Finally, we remark that the source data adopted by this work have been released as open-source at the link \cite{dataset}, to allow practitioners, industries, and academia to verify our claims and use them as a basis for further development.

\textcolor{black}{
\subsection{Experimental Performance Comparison}
\label{sec:perf_comp}
To provide further insights, in this section we compare the performance of \sol\ against the one of the most closed peer-reviewed scientific contribution at the time of this writing, that is the scheme proposed by the authors in~\cite{Alipour2019}.\\
The contribution in~\cite{Alipour2019} discusses a framework for the detection and brand identification of commercial drones based on Machine Learning, by using the same basic features adopted in our work, that are the packet size and packets interarrival times. However, despite \sol, the work in~\cite{Alipour2019} selects the one-vs-all logistic multi-class classification algorithm as the tool to discriminate drone traffic from the generic one. Unfortunately, the authors did make neither the raw data nor the code publicly available for any direct comparison. Therefore, following the discussion in the reference paper, we replicated the implementation of the method they proposed and we tested its performance. Specifically, we compared the overall Success Rate (SR) of \sol\ and the proposal in~\cite{Alipour2019} as for the drone detection performance in several real-life scenarios, matching the ones used in our analysis reported in Section~\ref{sec:detecting_a_drone}. \\
We investigated the performance of the approaches by increasing the number of samples used for the computation of the mean and standard deviation of the features, from 51 to 101 samples. We recall that, as depicted in Figure 7 of the manuscript, such a number of samples is directly connected to the detection delay, from a minimum value of $0.334$~s to the maximum value of $0.661$~s. The results are reported in Fig.~\ref{fig:comparison}.\\
\begin{figure}[htbp]
    \includegraphics[width=.9\columnwidth]{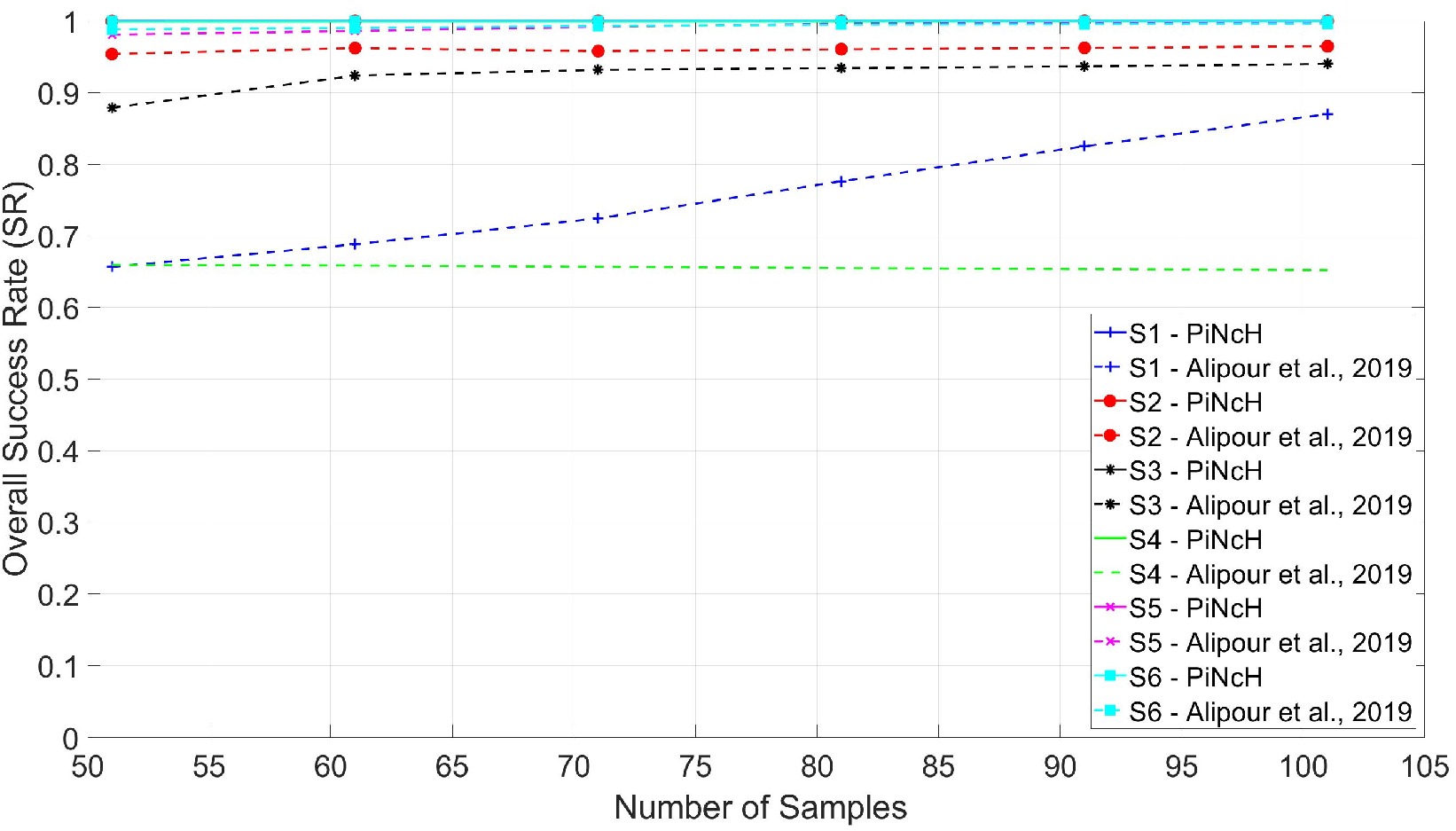}
    \centering
    \caption{Overall Success Rate (SR) in detecting the presence of a drone for the 6 different scenarios investigated in our paper, using \sol\ and the proposal by Alipour et al. in~\cite{Alipour2019}, by increasing the number of samples used for the computation of the features.}
    \label{fig:comparison}
\end{figure}
The results show that the performance and the overall success rate of the approach proposed by the authors in~\cite{Alipour2019} depends strictly on the selected scenario. While in the political meeting (S5) and local outdoor (S6) scenarios the performance of the benchmarking technique are similar to the ones of \sol\ (with detection rates that are in accordance to the ones reported in the reference paper), in the remaining four scenarios, the overall success rate is significantly lower than our technique. In addition, increasing the number of samples used for the computation of the statistical indexes of the features, i.e., increasing the detection delay, not always leads to a significant improvement.\\
Considering the same requirements on the detection delay, \sol\ is characterized by remarkable performance in the detection of a drone in all the 6 analyzed scenarios, 
and its performance always improve when increasing the detection delay. \\
Finally, despite the better performance with shorter detection delays of our proposed approach against the competing solutions, we stress that the major contributions of our work are the following: (i) we demonstrate that network traffic analysis can be considered as a valuable and meaningful tool to detect Remotely Piloted Aircraft Systems (RPAS); (ii) we experimentally show that network traffic analysis is a robust solution against packet losses and adversarial strategies; and, finally, that (iii) such technique could detect the presence of a drone, its current status, and its movements in a short time frame; all these features  being seamlessly integrable with both commercial devices and additional drone detection solutions.
}

\section{Conclusions and Future Work}
\label{sec:conclusions}
In this paper, we have introduced \sol: a methodology to detect the presence of a remotely-controlled drone in several heterogeneous environments with a high degree of assurance and a very short delay. \sol\ is also capable of identifying the drone's movements. These results are achieved without resorting to any active techniques, but just eavesdropping the radio traffic. 
In particular, we proved that network traffic classification can be effectively used to detect and identify the 3DR SOLO---the most popular open-source drone---as well as all the \acp{UAV} employing the popular operating system ArduCopter (such as some DJI and Hobbyking vehicles). Indeed, we provide an upper bound on the detection delay when using the aforementioned methodology.

We tested our methodology against six different scenarios and we proved that \sol\ can detect an RPAS drone in less than 0.28 seconds with a \ac{SR} of about 0.998 (worst case). Further, \sol\ can be effectively used to identify each of the drone's movements in about 1.5 seconds, with a \ac{SR} greater than 0.95. 
The comparison against the competing solution in the literature does show that \sol\ enjoys superior performance in several scenarios.
We also evaluated the effectiveness of \sol\ in an outdoor scenario, showing that our methodology is still quite robust also when more than 70 \% of the packets are lost. Finally, we also evaluated the robustness of \sol\ to evasion attacks, where the profile of the traffic of the drone is modified on purpose to avoid detection. In this scenario,  we showed that the effectiveness of such strategy is strongly dependent on the specific scenario, and likely not of general applicability, since evasion techniques could severely degrade the controller-drone channel, and hence its maneuverability.

Given that our study has been performed on a popular open-source operating system for drones---and all the collected data have been publicly released---, it can be also used to detect and identify different brands and models of drones, other than applying to other contexts as well.

\section*{Acknowledgements}
The authors would like to thank the anonymous reviewers for their  comments and insights, that have helped improving the quality of the paper.

This publication was partially supported by awards NPRP-S-11-0109-180242, UREP23-065-1-014, NPRP X-063-1-014, and GSRA6-1-0528-19046, from the QNRF-Qatar National Research Fund, a member of The Qatar Foundation. The information and views set out in this publication are those of the authors and do not necessarily reflect the official opinion of the QNRF.

\section*{References}
\balance
\bibliographystyle{IEEEtran}
\bibliography{conference}

\end{document}